\documentclass[apj,twocolumn,floatfix]{openjournal}

\usepackage{xcolor}
\usepackage{textgreek}
\usepackage[utf8]{inputenc}
\usepackage[english]{babel}
\usepackage{amsmath}	
\usepackage{amssymb}	
\usepackage{booktabs}
\usepackage[T1]{fontenc}
\usepackage{ae,aecompl}
\usepackage{graphicx}	
\usepackage{hyperref}
\usepackage{color,colortbl}
\definecolor{linkcolor}{rgb}{0.0,0.3,0.5}
\hypersetup{
    unicode, 
    colorlinks=true,
    linkcolor=linkcolor,
    citecolor=linkcolor,
    filecolor=linkcolor,
    urlcolor=linkcolor,
    breaklinks=true,
}
\usepackage{tensind}
\usepackage{cleveref}
\usepackage{multirow}
\usepackage{orcidlink}
\usepackage{pifont}
\usepackage{natbib}
\usepackage{subfigure}
\usepackage{soul}
\usepackage{enumitem}
\usepackage{rotating}
\usepackage{makebox}

\newcommand{\kms}{\,km\,s$^{-1}$}
\newcommand{\ergs}{\,erg\,s$^{-1}$}
\newcommand{\ang}{\,\AA}
\newcommand{\halpha}{H$\alpha$}
\newcommand{\hbeta}{H$\beta$}
\newcommand{\lam}{$\lambda$}
\newcommand{\llam}{$\lambda\lambda$}
\newcommand{\mum}{\,$\mu$m}
\newcommand{\othree}{[\ion{O}{3}]}

\begin{document}

\title{Kinematic Stratification in Extremely Red Quasars Revealed by JWST}
\shorttitle{Kinematic Stratification in ERQs}
\shortauthors{Neustadt et al.}



\author{\vspace{-1.4cm}
Jack~M.~M.~Neustadt\,\orcidlink{0000-0001-7351-2531}$^1$,
Nadia~L.~Zakamska\,\orcidlink{0000-0001-6100-6869}$^1$
Yu-Ching~Chen\,\orcidlink{0000-0002-9932-1298}$^1$
Andrey~Vayner\,\orcidlink{0000-0002-0710-3729}$^{2,3}$
Fred~Hamann$^4$
Marie~Wingyee~Lau\,\orcidlink{0000-0001-9755-9406}$^5$
Serena~Perrotta\,\orcidlink{0000-0002-2451-9160}$^6$
Kate~Rowlands\,\orcidlink{0000-0001-7883-8434}$^{1,7}$
Sylvain~Veilleux\,\orcidlink{0000-0002-3158-6820}$^8$
Dominika~Wylezalek\,\orcidlink{0000-0003-2212-6045}$^9$
\vspace{0.15cm}}

\affiliation{$^1$Department of Physics and Astronomy, Bloomberg Center, The Johns Hopkins University, Baltimore, MD 21218, USA\\
$^2$Florida Gulf Coast University, 10501 FGCU Blvd. South, Fort Myers, 33965, FL, USA\\
$^3$IPAC, California Institute of Technology, 1200 E. California Boulevard, Pasadena, CA 91125, USA\\
$^4$Department of Physics \& Astronomy, University of California, Riverside, Riverside, CA 92521, USA\\
$^5$Department of Physics and Astronomy, University of California, Riverside, CA 92521, USA\\
$^6$Department of Astronomy and Astrophysics, University of California, San Diego, La Jolla, CA 92093, USA\\
$^7$Department of Astronomy and Joint Space-Science Institute, University of Maryland, College Park, MD 20742, USA\\
$^8$Zentrum für Astronomie der Universität Heidelberg, Astronomisches Rechen-Institut, Mönchhofstr 12-14, D-69120 Heidelberg, Germany}

\email[Corresponding author: ]{jneustadt@jhu.edu}


\begin{abstract}
We analyze the spectra of the central nuclei of extremely red quasars (ERQs) observed as part of the JWST ERS Q3D program.  We focus on the complex kinematic structures of the prominent rest-frame optical emission lines.  Our modeling allows us to deblend the lines and separate the emission into distinct kinematic components that imply velocity- and density-stratified gas structures on a range of physical scales within the ERQs.  Supplementing the JWST data with archival data, we analyze the spectral energy distributions (SEDs) of the ERQs and find they are consistent with a significantly dust-obscured central source with a small amount of relatively-unobscured UV/optical flux that is scattered into our line-of-sight.  While the kinematics of the UV and optical emission lines largely agree, the UV lines are dominated by scattered light. In contrast, the optical emission-line ratios indicate a combination of scattered and obscured emission.  Our analysis focuses on one ERQ, J0834, because its distinct spectroscopic features allow the emission to be easily decomposed into separate kinematic components.
\end{abstract}


\maketitle

\section{Introduction}\label{sec:intro}

The correlation between the mass of a supermassive black hole (SMBH) in the center of a galaxy’s bulge and the bulge velocity dispersion -- the $M$--$\sigma$ relation (e.g., \citealt{kormendy95,ho08,kormendy13,mcconnell13}) -- suggests that the growth of galaxies and their SMBHs are related.  This is not an obvious relation, as SMBHs often only make up < 1\% (typically 0.1\%) of their host bulge’s mass.  The growth of these SMBHs must be primarily due to accretion (as opposed to mergers, see \citealt{soltan82,yu02,merloni08}).  When the accretion rate is sufficiently high, the system becomes an active galactic nucleus (AGN), which produces a tremendous amount of radiation, sometimes outshining its host galaxy (thus becoming a quasar).  This radiation and any associated jets power outflows of gas and dust which then act as feedback on the accretion process, halting the growth of the SMBH and, potentially, the star formation in its host galaxy. 

However, the mechanics of feedback and outflows are quite complicated.  Outflows can be produced by jet activity and radiative launching, and while radio-quiet AGNs with no obvious jets dominate the number counts (e.g., \citealt{begelman84}), it is not clear which process dominates the impact on galaxy evolution.  Radiative launching can be roughly split into two physical processes: UV continuum and emission line photon absorption \citep{proga00,proga04} and dust driving \citep{ishibashi15}.  It is not yet known which process dominates and how the outflows affect different spatial scales in the galaxy \citep{thompson05,murray05,faucher12,nims15}.

In order to study feedback as it is occurring, we can look at AGNs that are both luminous and heavily obscured: a luminous AGN should not be obscured for very long \citep{jun21}, as the radiation-driven outflows should ``blow out'' the obscuration (e.g., \citealt{hopkins06}).  Extremely red quasars (ERQs) are one example of luminous, obscured AGNs. ERQs mostly exist at cosmic noon ($z \sim$ 2--3) and are notable for having extremely red rest-frame optical-to-infrared (IR) colors \citep{ross15}, similar to dust-obscured star-forming galaxies \citep{eisenhardt13}.  They are often extremely luminous in the IR, up to $10^{46-47}$\ergs, making them some of the most luminous objects in the Universe \citep{hamann17}.  Based on the X-ray luminosities and column densities \citep{goulding18,ma24} and the near impossibility of achieving such bolometric luminosities with star formation (>10$^{3} \rm \,M_\odot\,yr^{-1}$), it has been proposed that these are near- or super-Eddington-accreting AGNs embedded in the dusty remnants of a recent merger \citep{zakamska19,perrotta19}. 

ERQs have peculiar spectral energy distributions (SEDs).  While they show evidence for significant dust obscuration, they also show relatively blue UV-optical continua with spectral slopes typical of AGNs.  The UV flux is also highly polarized \citep{alexandroff18,zakamska23}, which indicates that it is scattered into our line-of-sight by gas on scales comparable to or exceeding the scales of circumnuclear obscuration.  The SEDs of ERQs are not unlike Hot Dust Obscured Galaxies (HotDOGs, \citealt{eisenhardt13,assef15,jun20,vayner25}), where UV/optical photons are thought to be mostly absorbed by an IR-bright obscurer and escape by scattering, and ``little red dots'' (LRDs, e.g., \citealt{greene24,matthee24}), where the UV/optical continuum is uniquely dominated by ``v-shaped'' SED centered around the Balmer break (e.g., \citealt{wang25}).  The similarity with HotDOGs is very likely physical, in that the physical scenarios of the two classes of objects are likely analogous, whereas the similarity with LRDs is perhaps only observationally coincidental as the physical interpretations of LRDs are different (see e.g., \citealt{inayoshi25,liuh25,naidu25}).

ERQs are spectroscopically distinct from typical AGNs.  In the optical, the typical nuclear spectrum of a moderate-luminosity AGN has two ``types'' of lines -- narrow and broad.  The narrow lines include forbidden emission like \othree~\llam4959,5007, [\ion{O}{1}]~\llam6300,63, [\ion{N}{2}]~\llam6548,83, [\ion{S}{2}]~\llam6716,31, which can only be produced by collisional excitation in low-density gas, and permitted emission, like the Balmer series and \ion{He}{2}~\lam4686, which is predominantly produced by recombination.  The associated Narrow Line Region (NLR) is thought to consist of clouds of gas far away from the central accreting SMBH, up to several kpcs away, based on their relatively narrow velocity dispersions (full width at half maximum, FWHM < 1200\kms; see, e.g., \citealt{hao05}).  The broad lines, which are only permitted lines and have FWHMs >2000\kms, are associated with the Broad Line Region (BLR), which is thought to consist of clouds relatively near the central SMBH (light-weeks to light-months) with some unknown geometry (for an overview of the unification model, see \citealt{antonucci93}).  There is not always a clean divide between BLRs and NLRs, and indeed there is evidence for ``intermediate'' line regions (ILR, \citealt{brotherton94}) whose kinematics may be dominated by inflows at the outer regions of BLRs \citep{hu08} and whose ionization physics may be dominated by dust sublimation and higher densities \citep{adhikari16}.  Nevertheless, an emission line in a typical AGN spectrum will either be narrow, with no broad component (typical of a Type 2 AGN), or a combination of a narrow and a broad component (typical of a Type 1 AGN).  No such distinction can be made for ERQs: all of the emission lines are broad, including the forbidden lines \citep{zakamska16,perrotta19,lau24}, and many clearly have complicated velocity profiles \citep{gillette24}. In addition, ERQs also show evidence for high velocity, high velocity-dispersion, high-ionization outflows in the form of broad UV absorption lines (BALs) \citep{hamann17}, but BALs are not uncommon in AGNs (e.g., \citealt{ganguly08}).  

The focus of this study is an analysis of the James Webb Space Telescope (JWST) spectra of six ERQs obtained as part of the Q3D ERS program (ID: 1335, PI Wylezalek, Co-PIs: Veilleux, Zakamska, Software Lead: Rupke; ID: 2457: PI: Vayner).  These six ERQs were chosen based on the availability of archival data for these objects and the detectability of prominent rest-frame emission lines by NIRSpec.  So far, the publications analyzing these data \citep{vayner21a,wylezalek22,vayner23,gillette23,vayner24} have focused on the extended gas surround the ERQs, with one exception: \citet{bertemes25} examined the nuclear spectrum of J165202.64+172852.3 (hereafter, J1652) and pointed out that single-epoch spectra of ERQs give very conflicting measurements of the SMBH mass, which is unsurprising for a population with very different emission line profiles from typical AGNs.  Here we analyze the JWST spectra in concert with archival data to determine the kinematics and structure of the innermost regions of these ERQs.

In Section~\ref{sec:obs}, we discuss our sample of ERQs, the JWST observations, and the archival data.  In Section~\ref{sec:j0834}, we focus on the spectra of one ERQ of particular interest,  J083448.48+015921.1 (hereafter, J0834) for which our analysis is the most extensive.  In Section~\ref{sec:spec}, we repeat parts of our spectroscopic analysis from Section~\ref{sec:j0834} for the other ERQs.  In Section~\ref{sec:sed}, we analyze the SEDs of the ERQs.  In Section~\ref{sec:discuss}, we discuss our results and their physical interpretations, and in Section~\ref{sec:conc}, we summarize our findings and look to future projects.

We use rest-frame wavelengths, unless explicitly stated otherwise.  For the purposes of computing luminosity distances, $D_L$, we use redshifts and the \citet{wright06} calculator assuming $\Lambda$CDM, $h_0 = 69.6$, $\Omega_m = 0.286$, and a flat Universe. \\

\section{Observations}\label{sec:obs}

The physical parameters and available data for our ERQ sample are listed in Table~\ref{tab:sample}.  We use the redshifts from \citet{perrotta19} -- as we shall see in later Sections, there are offsets between the centroids of the emission line profiles and the rest-frame velocities implied by these redshifts, and so the ``true'' redshifts of these sources are likely different from the values in Table~\ref{tab:sample}.  

\begin{table*}
\centering
\caption{Physical parameters and data availability of ERQ sample}
\begin{tabular}{lccccc} \toprule
ERQ & SDSS Name & Redshift$^{a}$ ($z$) & X-Shooter & ALMA \\ \midrule \midrule
J0832 & J083200.20+161500.3 & 2.4252 & (1/3)$^{b}$ & ---  \\ 
J0834 & J083448.48+015921.1 & 2.5928 & \checkmark & \checkmark  \\ 
J1217 & J121704.70+023417.1 & 2.4266 & \checkmark  & \checkmark\\ 
J1232 & J123241.73+091209.3 & 2.3886 & \checkmark & \checkmark  \\ 
J1652 & J165202.64+172852.3 & 2.9482 & ---  & --- \\ 
J2215 & J221524.00\makebox*{+}[c]{--}005643.8 & 2.4975 & \checkmark & \checkmark \\ 
\bottomrule
\end{tabular}
\begin{flushleft}
\textit{Notes:} $^{a}$From \citet{perrotta19}.  $^{b}$Only includes the UV channel. 
\end{flushleft}
\label{tab:sample}
\end{table*}

\subsection{JWST}

The six targets were observed with JWST with the integral field unit (IFU) mode of the Near Infrared Spectrograph \citep[NIRSpec;][]{boker22} between November 2022 and June 2024. The field-of-view of the JWST IFU data is $3 \arcsec\times3\arcsec$. The G235H/F170LP grating (1.66--3.17\mum) was used for all targets, with spectral resolutions of $R\sim$ 1500--3500 ($\Delta v \sim$~85--200\kms). We obtained a leakage exposure with the same grating in one dither position for each target to remove the light from the failed-open shutters and the light leaking through the micro-shutter assembly (MSA, \citealt{deshpande18}). We used a nine-point small cycling dither pattern for the three targets in Program 1335, and a ten- or eleven-point small cycling dither pattern for the five targets in Program 2457 to improve the spatial sampling and help us more accurately characterize the point-spread function (PSF). 

Because the analysis of the kinematic profiles of the emission lines and the analysis of the SED require different data, we extract two spectra from our 3D NIRSpec datacubes using \textsc{q3dfit} \citep{rupke23}.  For our kinematics analysis, we extract spectra using a small aperture (typically a 3-spaxel radius) centered on the brightest spaxel of the data.  This way, we have a high signal-to-noise (S/N) for our emission lines, though the absolute fluxes are likely incorrect because the aperture is smaller than the PSF.  These spectra are analyzed in Sections~\ref{sec:j0834} and \ref{sec:spec}.  For our SED analysis, we extract spectra using a large enough aperture for the complicated PSF of the instrument to be negligible (typically 20--30-spaxel radius).  These spectra are analyzed in Section~\ref{sec:sed}.

\begin{figure*}
\centering
\includegraphics[width=\textwidth]{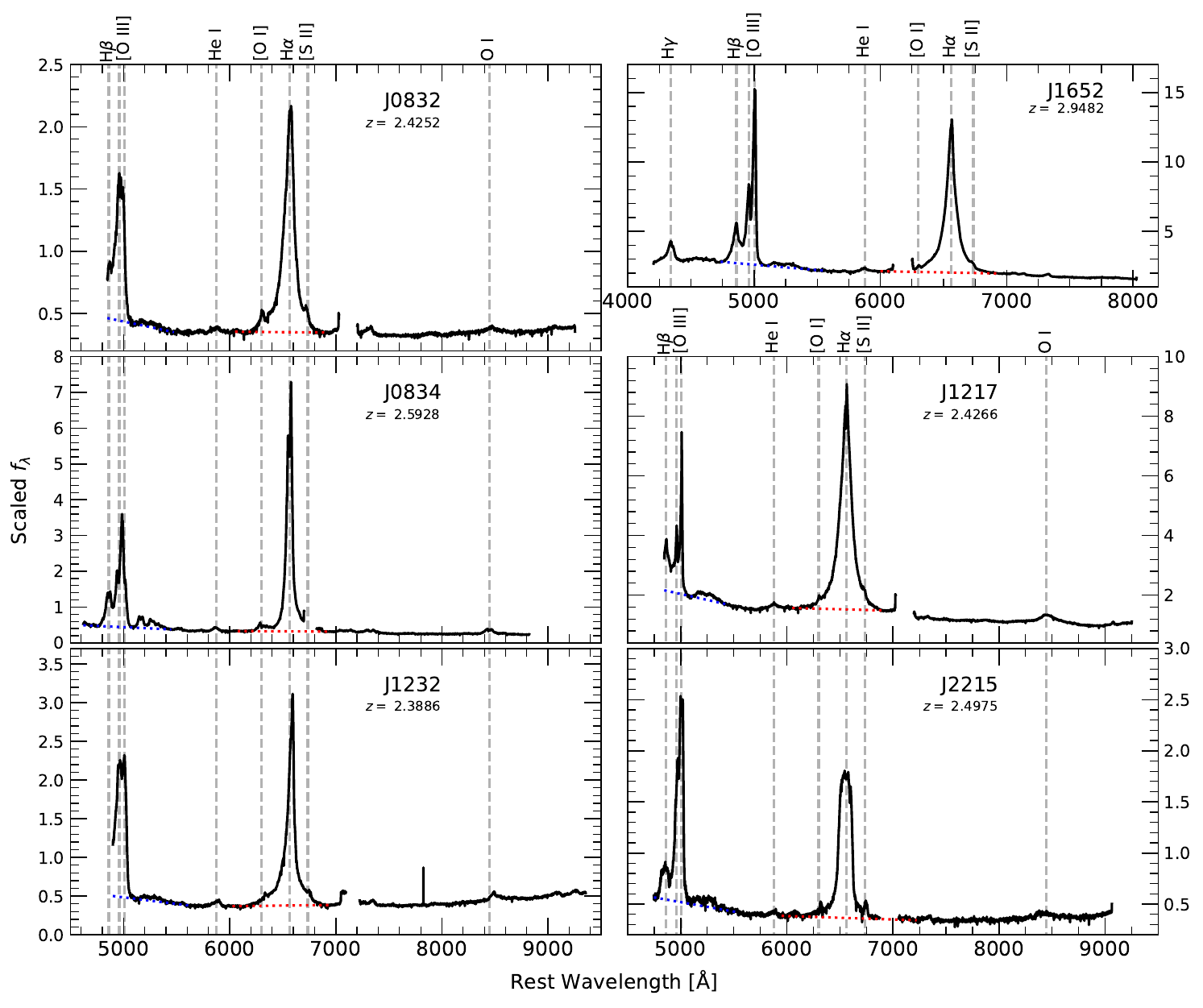}
\caption{JWST NIRSpec spectra of the ERQs. Prominent emission lines are labeled.  The gap in the spectra is due to the gap between the NRS1 and NRS2 channels. For analysis of spectral lines, we subtract out the continuum model shown by the dotted lines.}\label{fig:jwst}
\end{figure*}

\subsection{ALMA Data}

We use observed-frame 240\,GHz ALMA observations (ID: 2017.1.00478.S, PI: Hamann), corresponding to rest-frame 300--400\mum.  We extract fluxes using \textsc{imstat} in the Common Astronomy Software Applications package (CASA, \citealt{casa20}).  J1652 does not have associated ALMA data.

\subsection{Archival Data}

We include $ugriz$ photometry and spectra from the Sloan Digital Sky Survey (SDSS) and W1, W2, W3, W4 photometry from the Wide-field Infrared Survey Explorer (WISE, \citealt{wise}).  We also include archival spectra, and Table~\ref{tab:sample} shows the data availability for each object.   The X-Shooter spectra from the European Southern Observatory (ESO) Very Large Telescope (VLT) are obtained by \citet{perrotta19}.  In the observed-frame, these observations span 3000\ang\ to $\sim$2\mum, roughly corresponding to 1000--5000\ang\ in the ERQs' rest-frames.  For J1652, we include a spectrum from the Gemini Near-InfraRed Spectrograph (GNIRS) on Gemini North obtained by \citep{perrotta19} that spans 8500\ang\ to 2.5\mum\ in the observed-frame, roughly corresponding to 2000--6000\ang\ in the ERQ's rest-frame.  

We include observed-frame 6.2\,GHz observations of our ERQ sample obtained with the Karl G. Jansky Very Large Array (VLA) and 1.4\,GHz as measured by the Faint Images of the Radio Sky at Twenty Centimeters (FIRST, \citealt{becker95}) survey. The data are summarized in \citet{hwang18}. \\

\section{Spectroscopic Analysis of J0834}\label{sec:j0834}

We are interested in seeing if the JWST spectra show evidence for similar or distinct kinematics (1) when comparing the forbidden and permitted emission lines in the optical, and (2) when comparing UV and optical emission lines.  We focus on J0834 because of the distinct peaks in its Balmer and \othree\ emission profiles, which allow us to more easily deblend the lines into different kinematic components.

In this Section, we first explore the kinematic separation of the forbidden and permitted emission lines.  \othree~\llam4959,5007 are forbidden lines, which are collisionally de-excited if the density is too high ($n_{\rm cr}^{\rm} = 10^{5.8}~{\rm cm^{-3}}$), so they tend to originate on scales of hundreds of parsecs or more from the nucleus \citep{baskin05}.  Balmer lines are permitted and can originate in both the nuclear and the larger extended regions \citep{antonucci93,zakamska03,hao05}. 

Next, we explain the physical conditions of the lines.  In a typical AGN, the kinematic profiles of the optical and UV lines are similar because they originate from the same BLR (or NLR) clouds.  Dust complicates this because UV emission is both more scattered and more absorbed by dust than optical emission is.  Since ERQs are both heavily obscured and show evidence for scattered UV emission \citep{alexandroff18,zakamska23}, we will explore these effects by comparing the kinematics of the UV and optical lines.

\subsection{Modeling for J0834}

As seen in Figure~\ref{fig:jwst}, the emission lines cannot be cleanly broken down into ``narrow'' and ``broad'' components, and the \othree\ emission is often comparably broad to the Balmer lines.  Furthermore, line blending complicates line profile extraction.  To address this, we will model the following emission lines which we will term a spectral template: \\
$\bullet$ \halpha~\lam6563 \\
$\bullet$ [\ion{N}{2}]~\llam6548,83, with ratios fixed at $\sim$1:3 \\
$\bullet$ [\ion{O}{1}]~\llam6300,63, with ratios fixed at $\sim$3:1 \\ 
$\bullet$ [\ion{S}{2}]~\llam6716,31, with ratios allowed to vary as tracers for density. \\
$\bullet$ \hbeta~\lam4861 \\
$\bullet$ \othree~\llam4959,5007, with ratios fixed at $\sim$1:3 \\
$\bullet$ \ion{Fe}{2} complex defined using lines from \citet{kovacevic10}, with line ratios done assuming $T = 10000\,K$ and including lines from I~Zw~1. We allow for the ratios of different multiplets to vary, and we do not fix the widths of these lines to be the same as the other emission lines in the complex.

We model the data with multiple spectral templates, where each template has an independent velocity offset and velocity dispersion.  The ratio of integrated \halpha\ to \hbeta\ flux has a fixed lower limit of 2.86 as set by Case B recombination physics.  While the physics of BLRs are more complicated than simple recombination physics, we do not expect the Balmer decrement to be lower than the Case B limit.  For J0834, we need multiple spectral templates to include [\ion{O}{1}] and [\ion{N}{2}] to model the observed flux, but for other ERQs, we only need one spectral template to include [\ion{O}{1}] and [\ion{N}{2}].  For each velocity offset of the spectral templates, the offsets are determined relative to the redshift in Table~\ref{tab:sample}.  Therefore, the ``true'' rest-frame of the host galaxy may more accurately correspond to the velocity offset of a given spectral template.

While each spectral template has many emission line features, the focus of the spectral templates method -- and where this method is most robust -- is in separating the kinematics of the Balmer- and \othree-emitting gas.  From this we can make more informed analyses of other emission features, but there are still limits to this method.  There are multiple degeneracies between the free parameters of the templates, most notably with regards to the [\ion{N}{2}]/\halpha\ ratio of templates with large velocity dispersions, which in turn affects the measured Balmer decrement.    

We highlight J0834 because we see the same kinematic patterns in multiple emission lines.  We use a total of four spectral templates for modeling the emission lines.  The templates are as follows: \\
$\bullet$ broad blueshifted: Template 0834-A \\
$\bullet$ narrow redshifted: Template 0834-B \\ 
$\bullet$ narrow blueshifted: Template 0834-C \\ 
$\bullet$ very broad redshifted: Template 0834-D \\ 
The model parameters for template are each presented in Table~\ref{tab:j0834}.  The individual spectral templates as well as the composite fits are shown in Figure~\ref{fig:j0834-balmer}.  Two of the spectral templates, 0834-A and 0834-B, include \ion{Fe}{2} complexes. 

\begin{table*}
\centering
\caption{Modeled templates for J0834}
\begin{tabular}{lccccccc} \toprule
\multirow{2}{*}{Template} & Offset & FWHM & \multirow{2}{*}{\halpha/\hbeta} & \multirow{2}{*}{log \othree/\hbeta} & \multirow{2}{*}{log [\ion{N}{2}]/\halpha} & \multirow{2}{*}{log [\ion{O}{1}]/\halpha} & \ion{Fe}{2} FWHM \\ 
& [\kms] & [\kms] & & & & & [\kms] \\ \midrule \midrule
0834-A & --1290 $\pm$ 30\phantom{00} & 2320 $\pm$ 30\phantom{0} & 6.0 $\pm$ 1.3 & 1.0 $\pm$ 0.1  &  < --0.6\phantom{00} & --1.0 $\pm$ 0.1 & 1870 $\pm$ 130 \\ 
0834-B & 750 $\pm$ 10                & 890 $\pm$ 10             & 11.8 $\pm$ 0.9\phantom{0} & 0.0 $\pm$ 0.1  & --0.8 $\pm$ 0.1 &  --1.6 $\pm$ 0.1 & 1540 $\pm$ 80\phantom{0} \\ 
0834-C & --500 $\pm$ 10\phantom{0}   & 1230 $\pm$ 50\phantom{0} & 11.8 $\pm$ 1.0\phantom{0} & --0.7 $\pm$ 0.5\phantom{0} & --1.0 $\pm$ 0.4 & --1.4 $\pm$ 0.1 & ---\\ 
0834-D & 670 $\pm$ 60     & 7980 $\pm$ 130           & 3.3 $\pm$ 0.1 & --- &  ---  & --- & ---\\ \bottomrule
\end{tabular}
\begin{flushleft}
\textit{Notes:} Best-fit values and uncertainties are calculated using \textsc{scipy.optimize.curve\_fit} \citep{scipy}. 
\end{flushleft}
\label{tab:j0834}
\end{table*}

\subsection{Balmer lines and [O III]}\label{sec:j0834-balmer}

The strongest evidence for the necessity of the four templates comes from the Balmer and \othree\ profiles.  There are clear dual peaks in both the \halpha\ and \hbeta\ profiles, as well as in \othree, but the blue peaks for the Balmer and \othree\ emission lines are distinctly offset in velocity-space.  To show this, we group the templates into two pairs: one pair representing the blueshifted emission (Templates~0834-A and 0834-C, where the former has high \othree/\hbeta\ and the latter has little \othree), and one pair representing the redshifted emission (Templates~0834-B and 0834-D, where the latter has no associated forbidden line emission).  As seen in Table~\ref{tab:j0834}, the velocity offsets in each pair are comparable but -- crucially -- different.  

\begin{figure*}
\includegraphics[width=\linewidth]{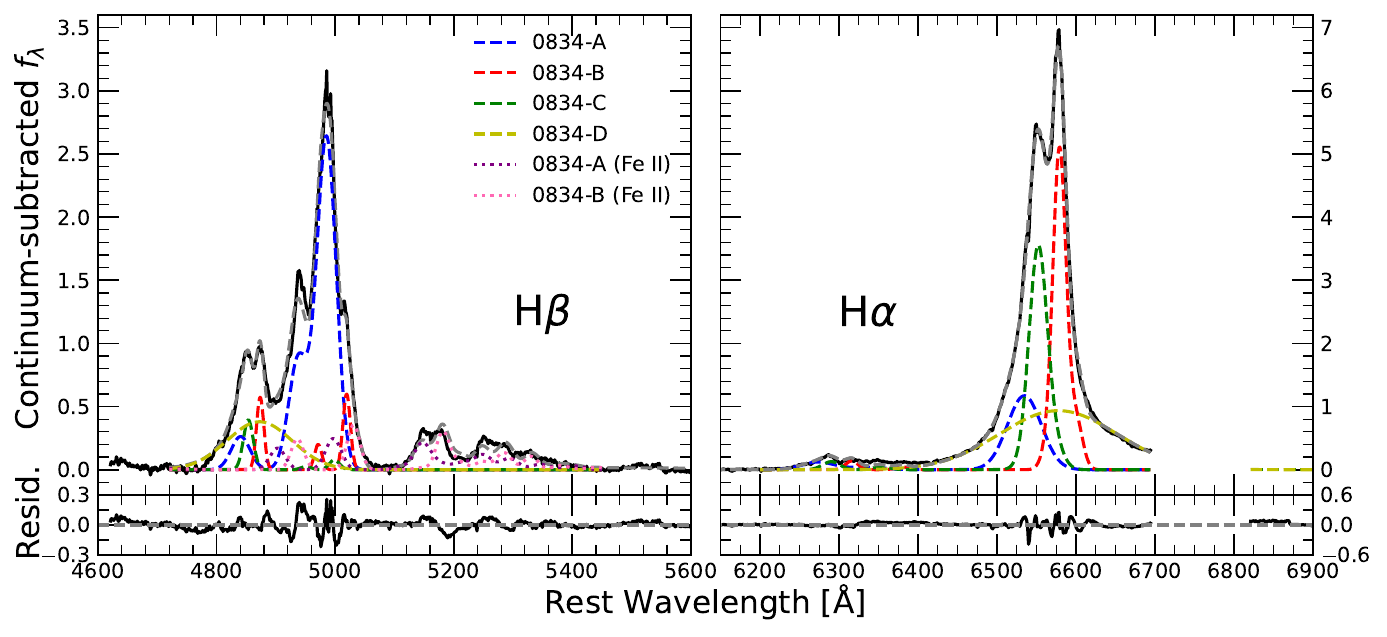}
\caption{Observed and modeled spectral profiles for the J0834 \hbeta\ + \othree\ (\textbf{left}) and  \halpha\ emission (\textbf{right}) along with residuals (\textbf{bottom}).  The colors correspond to the four templates from Tab.~\ref{tab:j0834}. }
\label{fig:j0834-balmer}
\end{figure*}

\begin{figure*}
\includegraphics[width=\linewidth]{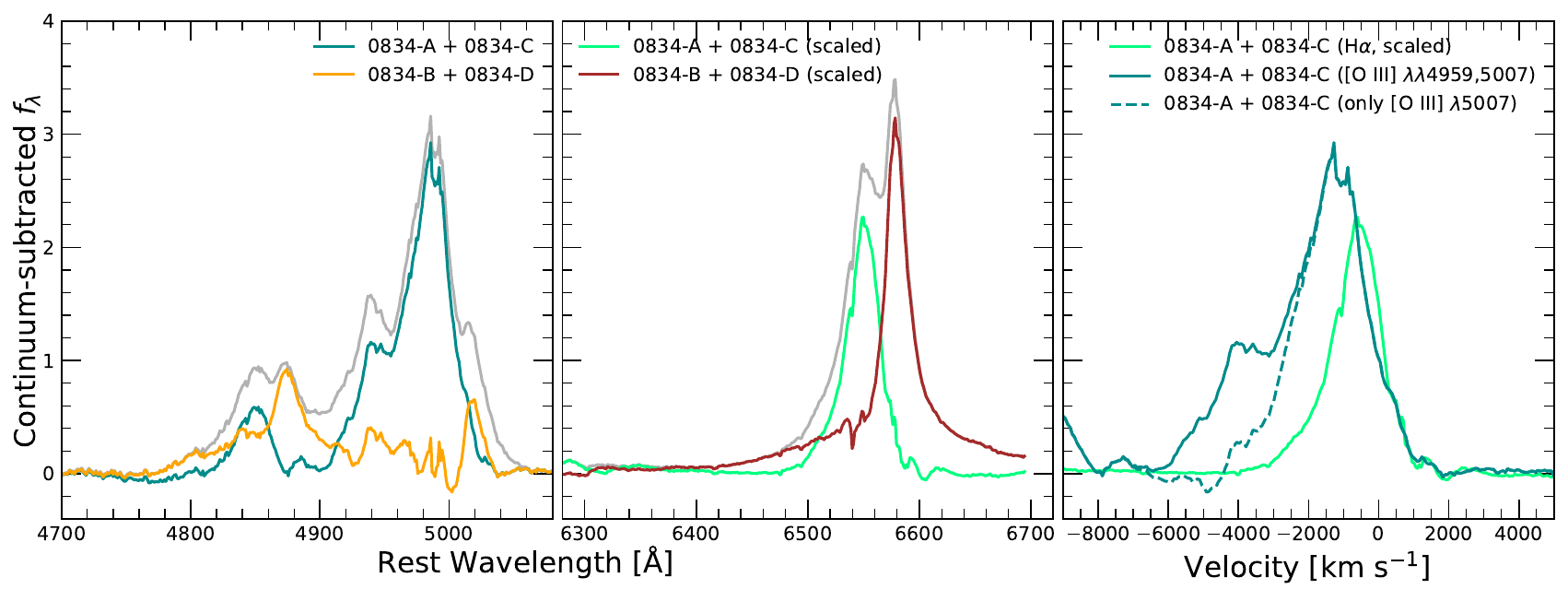}
\caption{ \textbf{Left:} The \hbeta\ and \othree\ emission profiles associated with Templates 0834-A and -C (cyan) and 0834-B and -D (orange).  The two profiles are constructed by subtracting the other templates from the total emission (see Sec.~\ref{sec:j0834-balmer}).  \textbf{Center:} Same as the previous panel, but for the \halpha\ and [\ion{N}{2}] emission profiles.  \textbf{Right:} The \halpha\ and \othree\ profiles of Templates 0834-A and -C.  In addition to showing the \othree\ doublet associated with Templates 0834-A and -C, we further highlight the distinct kinematics of the \othree~\lam5007 by subtracting the modeled emission of \othree~\lam4959.}
\label{fig:j0834-balmer-split}
\end{figure*}

We then subtract each pair of spectral templates from the total emission to highlight the emission being produced by the other two templates -- i.e., we subtract the 0834-A and 0834-C profiles from the total \halpha\ emission so that we can see the emission that corresponds to the 0834-B and 0834-D profiles -- with the results shown in Figure~\ref{fig:j0834-balmer-split}. We see that the bulk of the blueshifted \halpha\ and \hbeta emission is associated with Template 0834-C, whereas the bulk of the blueshifted \othree\ emission is associated with 0834-A.  In other words, the bulk of the blueshifted \othree\ emission is not being produced by the same gas producing the bulk of the blueshifted Balmer emission because Templates~0834-A and 0834-C correspond to two kinematically distinct regions. 

We are interested in seeing if J0834 has a BLR typical of AGNs -- that is, a broad and high-density kinematic component that consists of only permitted lines.  From Figure~\ref{fig:j0834-balmer} and \ref{fig:j0834-balmer-split}, we can see a broad base to the \halpha\ emission modeled by Template 0834-D that includes only Balmer emission.  The velocity offset of this template is roughly the same as the offset of 0834-B (see Tab.~\ref{tab:j0834}), but the widths are clearly different -- $\sim$1200\kms\ (0834-B) and $\sim$8000\kms\ (0834-D).  

While the sum of the four templates reproduces both \halpha\ and \hbeta\ well, it is clear that the fit overestimates the flux in the bluest edges of \hbeta\ (see Fig~\ref{fig:j0834-balmer} around 4750\ang).  When we attempted to fit the \halpha\ and \hbeta\ profiles separately, we found the same problem -- this is because the model is trying to produce the additional flux between the reddest component of \hbeta\ and the bluest wing of \othree~\lam4959, and it is likely due to contamination from unmodeled \ion{Fe}{2} multiplets or other lines. 

With the lack of associated forbidden emission, Template 0834-D appears to correspond to the emission associated with a typical AGN BLR.  Since 0834-B has a very similar velocity offset, 0834-B and 0834-D likely correspond to the emission from the AGN and its host (or rather, the AGN and its associated NLR), while 0834-A and 0834-C correspond to emission from kinematically distinct outflowing gases on larger scales.  

We are interested in comparing the obscuration of the different gas structures by measuring the extinction of the different spectral templates.  We can do this by comparing the observed Balmer decrements and the lower limit of 2.86 from Case B recombination.  We calculate extinction $A_V$ as
\begin{equation*}
    A_V = \frac{2.5}{A_\lambda({\rm H\beta})-A_\lambda({\rm H\alpha})} \, \log_{10} \Biggl[ {\frac{(\rm{H\alpha/H\beta})_{obs}}{2.86}} \Biggr] ~\rm mag
\end{equation*}
where (\halpha/\hbeta)$_{\rm obs}$ is the observed Balmer decrement and $A_\lambda$ is the extinction at wavelength $\lambda$.  The value left of the logarithm is dependent on the extinction curve and $R_V$ used, ranging from roughly 6.8--9.8 if we consider extinction curves of \citet{cardelli89} and \citet{gaskell07} and $R_V=3.1$--5.  For now, we are interested in the minimum required extinction to get the observed Balmer decrements.  This then yields $A_V \sim 2.1$\,mag for 0834-A, $A_V \sim 4.2$\,mag for 0834-B and 0834-C, and $A_V \sim 0.4$\,mag for 0834-D.  The two $A_V$ values of 0834-B and 0834-C are interesting, in that this would cause the optical-to-IR flux ratio to drop by a factor of $\sim$100, which is comparable to the difference of optical-to-IR flux ratios between the typical AGN and the ERQs \citep{hamann17}.    

At face value, it is quite surprising to find the smallest extinction for the component we hypothesize to be the most compact -- 0834-D.  But it is already well known from spectropolarimetric observations \citep{alexandroff18} that much, if not all UV continuum and broad lines of ERQs are not observed directly but are due to scattering.  While we do not have spectropolarimetric data to confirm this hypothesis directly, the relatively low value of extinction of this most-compact component can be due to the scattering.  The higher extinction values of the narrower, less compact lines could be due to intrinsic extinction or line-of-sight obscuration -- we shall investigate this in Section~\ref{sec:j0834-oi}.  We must also take into account that Template 0834-D overestimates the amount of \hbeta\ flux, leading to an underestimate of the Balmer decrement and thus the obscuration.  If we reduced the \hbeta\ contribution from 0834-D, this would increase the amount of \hbeta\ flux associated with the other templates, decrease the Balmer decrements, and thus reduce the implied obscuration.  

Furthermore, the Balmer decrement can be different if the emission is from sources other than recombination, like collisional excitation.  Indeed, collisional excitation is used to explain high Balmer decrements seen in LRDs \citep{burke25}, but these are sources with little dust extinction.  Because we already know ERQs are extremely dusty, invoking dust to explain the large Balmer decrements is more appropriate.

Overall, we highlight that Figure~\ref{fig:j0834-balmer-split} illustrates that, although the \othree\ is comparably broad to \hbeta\ in J0834 (and, as seen in Fig.~\ref{fig:jwst} and Sec.~\ref{sec:spec}, in other ERQs as well), our method of using multiple spectral templates allows us to see clear separation between emission from outflows and emission that is typical of NLRs and BLRs of moderate-luminosity AGNs.

\subsection{O I and its implications}\label{sec:j0834-oi}

We are interested in seeing if the kinematics of other emission lines match that of the Balmer and \othree\ lines.  We are especially interested in \ion{O}{1} because it can be a tracer for high-density gas.  \ion{O}{1}~\lam1304 and \lam8446 have emerged as lines of particular interest in some models of LRDs \citep{inayoshi25,tripodi25}.  In dense gas, \halpha\ and Ly$\beta$ can become optically thick, leading to Ly$\beta$ photons to be fluorescently scattered by neutral oxygen \citep{kwan81}.  The \ion{O}{1} photons from Ly$\beta$ fluorescence are distinct from recombination in that only \ion{O}{1}~\lam1304 and \lam8446 (and \lam11287) are produced, whereas recombination produces other lines like \ion{O}{1}~\lam7774 \citep{matsuoka07,matsuoka08}. Thus, strong \ion{O}{1}~\lam1304 and \lam8446 without comparably strong \lam7774 is a tracer for dense gas.  In contrast, [\ion{O}{1}]~\llam6300,63 is a tracer for radiative shocks propagating into quasi-neutral media.  

\begin{figure*}
\includegraphics[width=\linewidth]{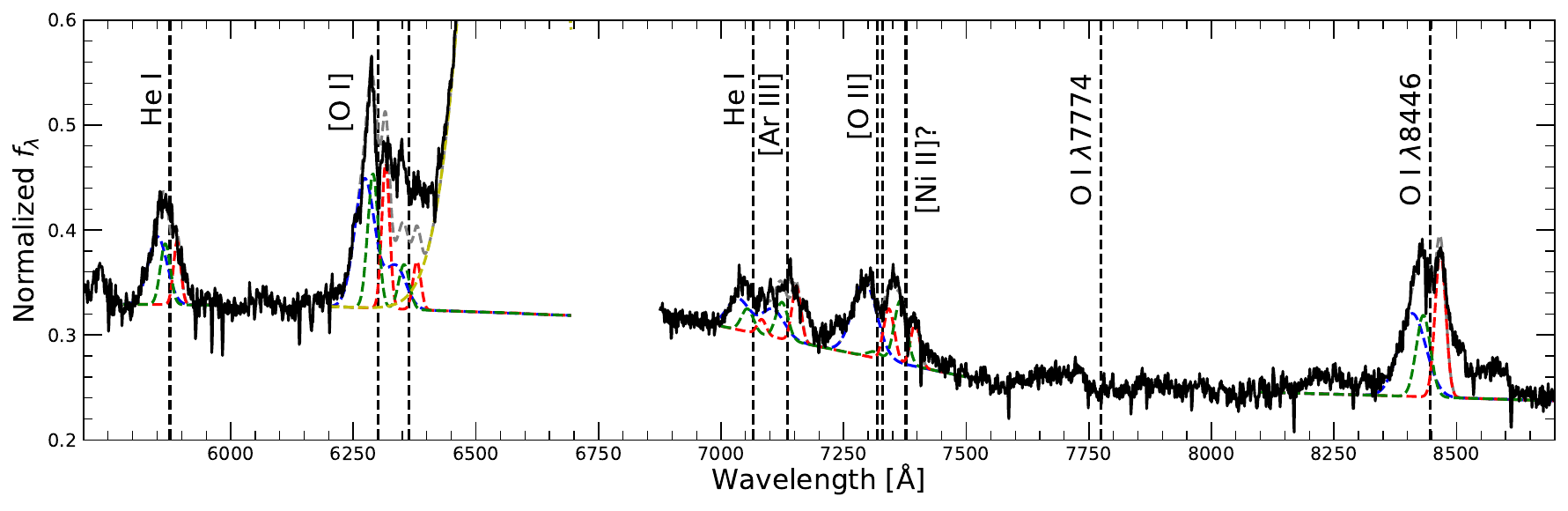}
\caption{Red optical and NIR spectrum of J0834.  We use the widths and offsets of Templates~0834-A, 0834-B, and 0834-C to fit the emission profiles of \ion{He}{1}~\lam5876, [\ion{O}{1}]~\llam6300,64, \ion{O}{1}~\lam8446, as well as a group of miscellaneous identified lines: \ion{He}{1}~\lam7065, [\ion{Ar}{3}]~\lam7136, [\ion{O}{2}]~\llam7320,31, and [\ion{Ni}{2}]~\lam7378 (the identification of the [\ion{N}{2}] line is uncertain).  We highlight the lack of detected \ion{O}{1}~\lam7774 emission.  The colors are the same as in Fig.~\ref{fig:j0834-balmer}.}
\label{fig:j0834-other}
\end{figure*}

We use the velocities and widths of the first three (0834-A, -B, -C) templates to fit the corresponding profiles for \ion{He}{1}~\lam5876, the [\ion{O}{1}]~\llam6300,64 doublet, \ion{O}{1}~\lam8446, and a group of lines in the 7000--7500\ang\ region shown in Figure~\ref{fig:j0834-other}.  There are some difficulties in matching the composite profiles of the various [\ion{O}{1}] doublet components and the blue wings of the broad \halpha\ line, most likely coming from very broad \halpha\ emission that may not be fully modeled by Template 0834-D. 
Otherwise, most of the \ion{He}{1}, \ion{O}{1}, and the emission between 7000--7500\ang\ is well-modeled by the three templates.  \ion{He}{1} and \ion{O}{1} show evidence for additional emission redward of the redshifted Template 0834-B, but it is unclear if the Balmer lines show similar emission.  

\begin{figure}
\includegraphics[width=\linewidth]{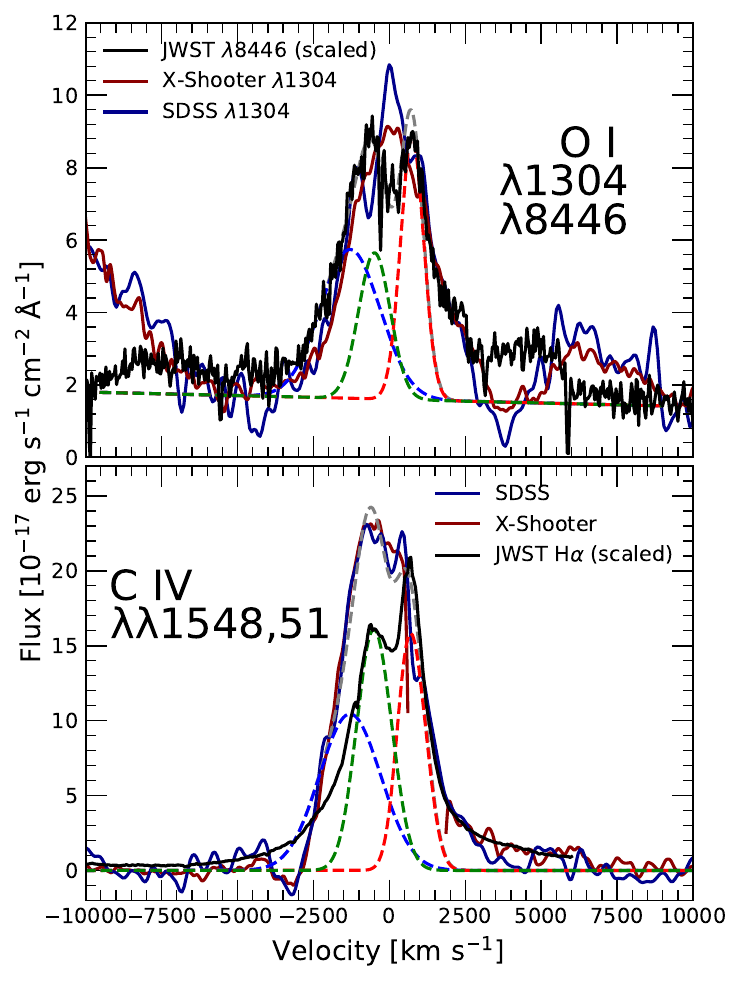}
\caption{\textbf{Top:} Velocity profiles of the \ion{O}{1}~\lam8446 line from the JWST spectrum and the \ion{O}{1}~\lam1304 line from the SDSS and X-Shooter spectra. The JWST spectrum is scaled to match the flux of the UV spectra with an observed \ion{O}{1}~\lam1304:\lam8446 flux ratio of 1:3--5, depending on the scaling used.  We also include the template fits to the \ion{O}{1}~\lam8446 profile that are also shown in Fig.~\ref{fig:j0834-other}.
\textbf{Bottom:} Velocity profiles of the \ion{C}{4}~\llam1548,51 line from the X-Shooter and SDSS spectra and the \halpha\ line from the JWST spectrum.  We include fits to the \ion{C}{4} line using the offsets and widths of templates 0834-A, -B, and -C.  In both panels, the SDSS and X-Shooter spectra are smoothed by Gaussian kernels with FWHMs of 1\ang\ to aid in visual comparisons.}
\label{fig:j0834-oi}
\end{figure}

We take a closer look at \ion{O}{1}~\lam1304 and \lam8446 in Figure~\ref{fig:j0834-oi}.  The velocity profiles are nearly identical, though there are slight differences -- the \lam8446 line shows evidence for two peaks corresponding to the offsets of templates 0834-B and 0834-C, but for \lam1304, this gap is ``filled in,'' likely by \ion{Si}{2}~\lam1307.   

While we observe both \ion{O}{1}~\lam1304 and~\lam8446, it is clear from Figure~\ref{fig:j0834-other} that there is little to no evidence for \ion{O}{1}~\lam7774, indicating that the \ion{O}{1} emission is likely due to Ly$\beta$ fluorescence.  In Ly$\beta$ fluorescence, the intrinsic flux ratio of \ion{O}{1}~\lam1304:\lam8446 is $\sim$6.5 based on a photon count ratio of 1:1 \citep{matsuoka07,matsuoka08}.  We can therefore also use the observed ratio as a measure of extinction.  The observed flux ratio of $\sim$3--5 (depending on the correction factor used between the JWST and UV spectra) can imply a range of extinction values, with a maximum of $A_V \sim$ 0.5~mag assuming $R_V = 5$ and a \citet{gaskell07} extinction curve. 

This value is much lower than the {\it minimum} extinction values implied by the Balmer decrements of Templates 0834-A, 0834-B, and 0834-C, yet we know that the kinematics of the Balmer lines are not dramatically different from those of \ion{O}{1} lines.  This means that the extinction implied by the Balmer decrement must be due to line-of-sight obscuration rather than intrinsic extinction from dust mixed in with the emitting gas  -- otherwise, the \ion{O}{1} line ratios would also yield a comparably large extinction value.  Indeed, the low extinction value of implied by the \ion{O}{1} line ratio, combined with the fact that Ly$\beta$ fluorescence traces dense gas, means that the emitting gases -- both the outflows traced by 0834-A and 0834-C and the NLR-like gas traced by 0834-B -- are relatively dust-free. 

Comparing Figures~\ref{fig:j0834-balmer-split} and \ref{fig:j0834-oi}, it is clear that 0834-A produces most of the \othree\ whereas 0834-B and 0834-C produce most of the \ion{O}{1}.  This is evidence of kinematic stratification, where the broader Template 0834-A is the lowest density, highest velocity, and perhaps largest-scale outflow, whereas the narrower Template 0834-B and 0834-C are lower velocity and higher density and, thus, likely closer in.  In a typical Seyfert, the stratification goes the other way -- the high-density BLR gas has a higher velocity dispersion than the low-density NLR gas.  

In Figure~\ref{fig:j0834-oi}, we also show the kinematics of the \ion{C}{4}~\llam1548,51 emission line.  We fit the kinematic profile with three components that correspond to Templates 0834-A, -B, and -C, and we find that the profile is well-fit with these components.   This means that the kinematics of the \ion{C}{4} line, and as well as most of the UV lines, match with that of the optical emission lines.  The exception to this is that the \ion{C}{4} profile, along with the \ion{S}{4} (see Fig.~\ref{fig:j0834-xs}) shows evidence for blue-shifted BALs.  This is evidence for an additional component to the outflowing gas that must be high-velocity, highly-ionized, and exterior to the gas corresponding to Template 0834-A.  

\subsection{Fe II modeling}\label{sec:j0834-feii}

We are interested in the kinematics of the \ion{Fe}{2} emission because it is thought to originate from a different region than the BLR in a typical AGN.  This is evidenced by the kinematic differences between \ion{Fe}{2} and \hbeta, and thus the \ion{Fe}{2} emission is linked to the ILR that may be dominated by inflowing or outflowing gas \citep{hu08,kovacevic10}.  Strong \ion{Fe}{2} is also linked to high Eddington ratios \citep{boroson92,kovacevic10}.

\begin{figure}
\includegraphics[width=\linewidth]{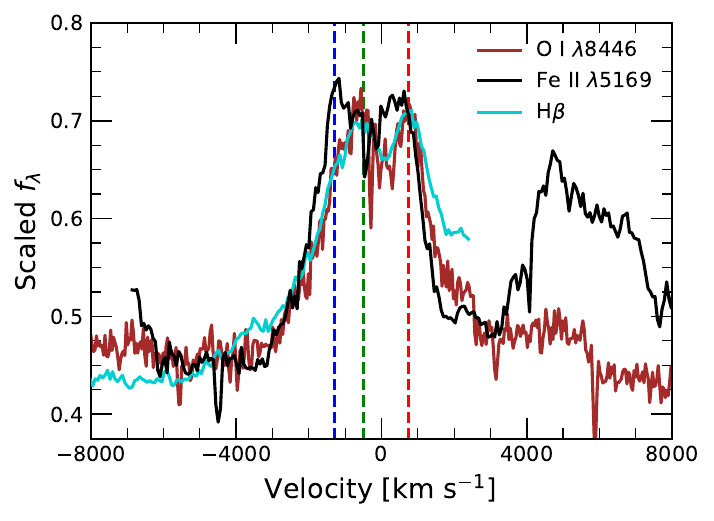}
\caption{Kinematic profiles of \ion{Fe}{2}~\lam5169 compared to \ion{O}{1}~\lam8446 and \hbeta.  The profiles largely agree with each other, albeit the peaks of the \ion{Fe}{2} emission are slightly offset from the other two lines.  This is strong evidence for two \ion{Fe}{2} kinematic components -- one linked to Template 0834-A and one from -B.}
\label{fig:j0834-feii-single}
\end{figure}

We see strong evidence that the \ion{Fe}{2} emission in J0834 displays two kinematically distinct components, most prominently in the \ion{Fe}{2}~\lam5169 feature.  The two peaks would then correspond to the blueshifted 0834-A and the redshifted 0834-B Templates.  We probe this model by fixing two \ion{Fe}{2} complexes to have the velocities of the Templates 0834-A and 0834-B, but allow the widths to be independent.  The resulting fits are shown in Figure~\ref{fig:j0834-balmer}.  We also highlight the double-peaked emission of the \ion{Fe}{2}~\lam5169 feature in Figure~\ref{fig:j0834-feii-single}.  When we compare this feature to \hbeta\ and \ion{O}{1}~\lam8446, the strong similarities between the kinematic profiles are very apparent, especially to the \ion{O}{1} line.  This aligns with the findings of studies of other AGNs, where \ion{Fe}{2} and Ly$\beta$-fluorescent \ion{O}{1} are thought to emerge from the same gas due to their usually-correlated kinematics in observations as well as their similar ionization potentials \citep{rodriguez02,matsuoka07,matsuoka08}.  

\begin{figure*}
\centering
\includegraphics[width=0.99\linewidth]{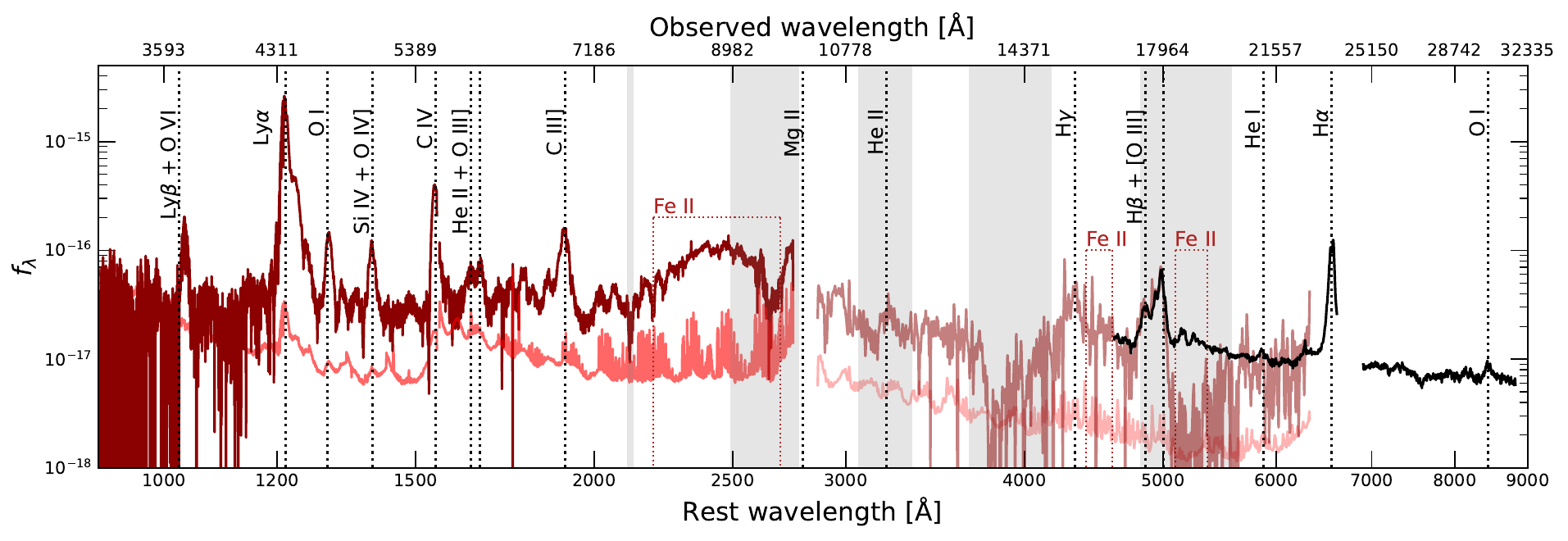}
\caption{Smoothed X-Shooter spectra (dark red) and error spectra (red) combined with JWST spectrum (black) for J0834.  We highlight the prominent UV bump created by \ion{Fe}{2} emission.  We also mark out telluric features (gray filled) that are relevant for the ground-based X-Shooter spectra.}
\label{fig:j0834-xs}
\end{figure*}

In Figure~\ref{fig:j0834-xs}, we show the JWST spectrum alongside the X-Shooter spectra, highlighting the presence of strong UV \ion{Fe}{2} emission relative to optical \ion{Fe}{2}.   Strong UV \ion{Fe}{2} relative to optical \ion{Fe}{2} can be indicative of resonantly scattered emission from outflows \citep{wang16}, which aligns with the evidence for Ly$\beta$ fluorescence.  We are unfortunately unable to separate the UV \ion{Fe}{2} complex into distinct kinematic components. \\

\section{Spectroscopic Analysis of the ERQ Sample}\label{sec:spec}

Next, we characterize the spectra of the other ERQs in our sample.  We do not go into as much detail as we do for J0834 in Section~\ref{sec:j0834}, focusing on the most important features of each ERQ.  Table~\ref{tab:erq_templates} summarizes the parameters of the templates discussed in this section.  In general, the templates are named such that `A' models the most blueshifted emission, `B' models the most redshifted, `C' models the intermediate velocity offset and width, and `D' models the broadest spectral features.  We include \ion{Fe}{2} complexes, but because the emission is much fainter than what we see with J0834, we do not include the widths or offsets in Table~\ref{tab:erq_templates}.

\begin{table*}
\centering
\caption{Modeled templates for other ERQs}
\begin{tabular}{lcccccccc} \toprule
ERQ & Template & Offset & FWHM  & \halpha/\hbeta\ & log \othree/\hbeta\ & log [\ion{N}{2}]/\halpha\ & log [\ion{O}{1}]/\halpha\ & log [\ion{S}{2}]/\halpha\ \\
& & [\kms] & [\kms] \\ \midrule \midrule
J0832 & 0832-A  & --1670 $\pm$ 40\phantom{00} & 4050 $\pm$ 90\phantom{0} & --- & --- & --- & --- & --- \\ 
      & 0832-B  &   450 $\pm$ 40 & 2420 $\pm$ 70\phantom{0} &  5.8 $\pm$ 0.8  & --0.7 $\pm$ 0.4\phantom{0} & 0.0 $\pm$ 0.1 & --0.4 $\pm$ 0.1\phantom{0} & --1.1 $\pm$ 0.1\phantom{0}  \\ 
       & 0832-D & --230 $\pm$ 50\phantom{0} & 9900 $\pm$ 180 & 2.9 $\pm$ 0.2 & --- & --- & --- & --- \\ \midrule
J1217 & 1217-A & --540 $\pm$ 30\phantom{0}   &  1950 $\pm$ 30\phantom{0} & 7.1 $\pm$ 2.8  & 1.3 $\pm$ 0.2 & --0.5 $\pm$ 0.3\phantom{0} & --- & ---\\
      & 1217-B &   180 $\pm$ 10   &           650 $\pm$ 10   &  3.2 $\pm$ 0.3  & 0.9 $\pm$ 0.1 & --0.3 $\pm$ 0.1\phantom{0} & --0.7 $\pm$ 0.1\phantom{0} & --0.8 $\pm$ 0.1\phantom{0}  \\ 
      & 1217-C & --180 $\pm$ 10\phantom{0}   &  5180 $\pm$ 70\phantom{0}  & 12.0 $\pm$ 1.5\phantom{0}  & --- & --- & --- & --- \\ 
      & 1217-D & --800 $\pm$ 40\phantom{0} & 13110 $\pm$ 160\phantom{0} & 2.9 $\pm$ 0.1  & --- & --- & --- & ---\\ \midrule 
J1232 & 1232-A & --2280 $\pm$ 70\phantom{00} & 4580 $\pm$ 20 & --- & --- & --- & --- & --- \\ 
      & 1232-B & 1400 $\pm$ 10\phantom{0}   & 970 $\pm$ 10 & --- & --- & --0.7 $\pm$ 0.1 & --1.1 $\pm$ 0.1\phantom{0} & --1.5 $\pm$ 0.1\phantom{0} \\ 
      & 1232-C & 280 $\pm$ 20    & 2190 $\pm$ 40 & --- & --- & --- & --- & --- \\ 
       & 1232-D & 300 $\pm$ 80   & 11790 $\pm$ 160 & 2.9 $\pm$ 1.7 & --- & --- & --- & --- \\ \midrule
J1652 & 1652-A &   --810 $\pm$ 20\phantom{0} &  2110 $\pm$ 20\phantom{0}  &  2.9 $\pm$ 0.8  & 1.2 $\pm$ 0.1 & 0.6 $\pm$ 0.1 & < --1.1\phantom{000} &  < --1.3\phantom{000} \\ 
      & 1652-B &   --20 $\pm$ 10 &     940 $\pm$ 10   & 2.9 $\pm$ 0.3  & 1.1 $\pm$ 0.1 & < --1.1\phantom{000} &  --1.2 $\pm$ 0.2\phantom{0}  & --1.4 $\pm$ 0.4\phantom{0} \\ 
      & 1652-C &    190 $\pm$ 30 &    6010 $\pm$ 130  & 2.9 $\pm$ 0.1 & --- & --- & --- & --- \\ 
      & 1652-D &  --1860 $\pm$ 240\phantom{0} & 13870 $\pm$ 540\phantom{0} & ---           & --- & --- & --- & --- \\ \midrule
J2215 & 2215-A & --2330 $\pm$ 50\phantom{00} & 1940 $\pm$ 100  & 3.3 $\pm$ 0.3  & 0.5 $\pm$ 0.1 & --0.2 $\pm$ 0.1\phantom{0} & --- & --- \\
      & 2215-B &    840 $\pm$ 10 & 970 $\pm$ 20 &    4.5 $\pm$ 0.7  & 0.9 $\pm$ 0.1 &  0.2 $\pm$ 0.1  & --0.5 $\pm$ 0.1\phantom{0} & --0.8 $\pm$ 0.1\phantom{0} \\ 
      & 2215-C & --360 $\pm$ 20\phantom{0}  & 1850 $\pm$ 50\phantom{0}  &  2.9 $\pm$ 0.4  & 0.9 $\pm$ 0.1 & < --0.8\phantom{000} & --- & ---\\ 
      & 2215-D &  --700 $\pm$ 40\phantom{0} & 7790 $\pm$ 140 &  8.9 $\pm$ 1.0  & --- & --- & --- & --- \\ 
\bottomrule
\end{tabular}
\begin{flushleft}
\textit{Notes:}  Best-fit values and uncertainties are calculated using \textsc{scipy.optimize.curve\_fit}.  Because of the blue cut-off of \hbeta\ in the spectra J0832 and J1232, there are large degeneracies between model parameters, and so the values of the Balmer decrement, [\ion{O}{3}]/\hbeta, and [\ion{N}{2}]/\halpha\ for Templates 0832-A, 1232-A, 1232-B ([\ion{N}{2}]/\halpha\ is reported), and 1232-C are not reported.  
\end{flushleft}
\label{tab:erq_templates}
\end{table*}


\begin{sidewaysfigure*}
\vspace{-10cm}
\centering
\subfigure{
\includegraphics[width=0.45\textwidth]{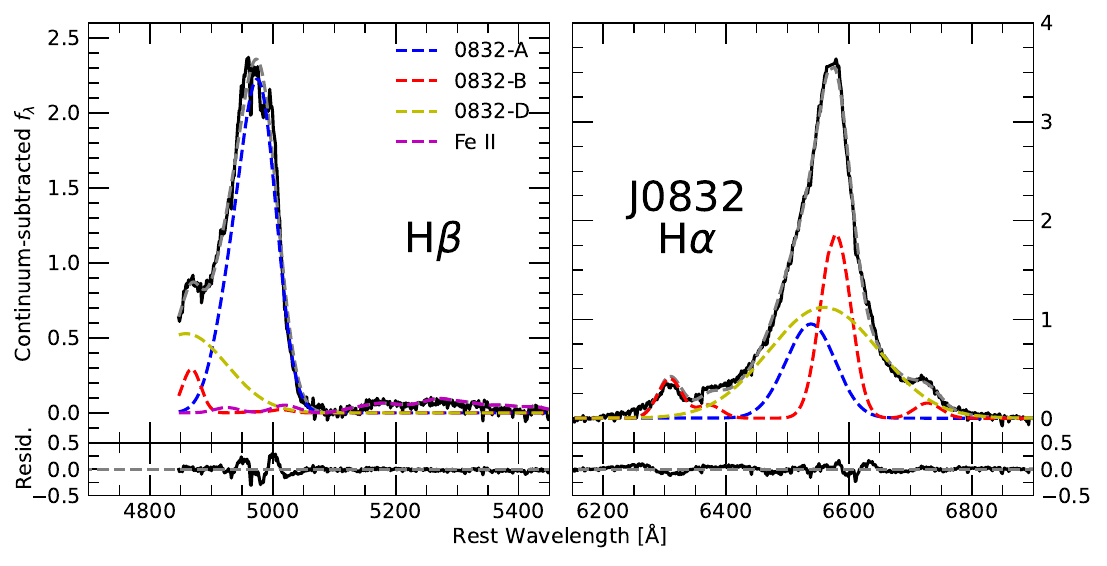}}
\subfigure{
\includegraphics[width=0.45\textwidth]{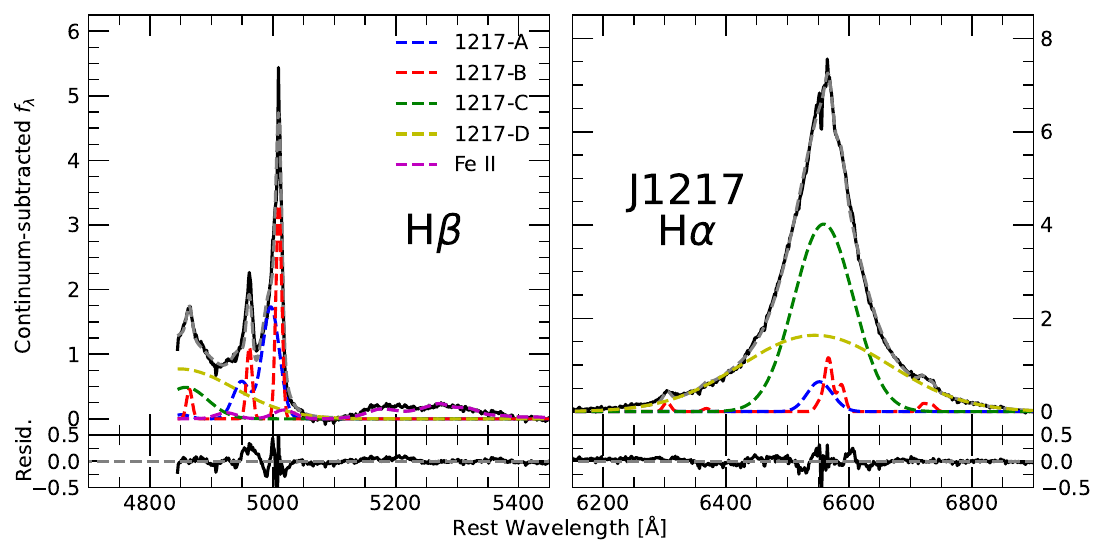}}
\subfigure{
\includegraphics[width=0.45\textwidth]{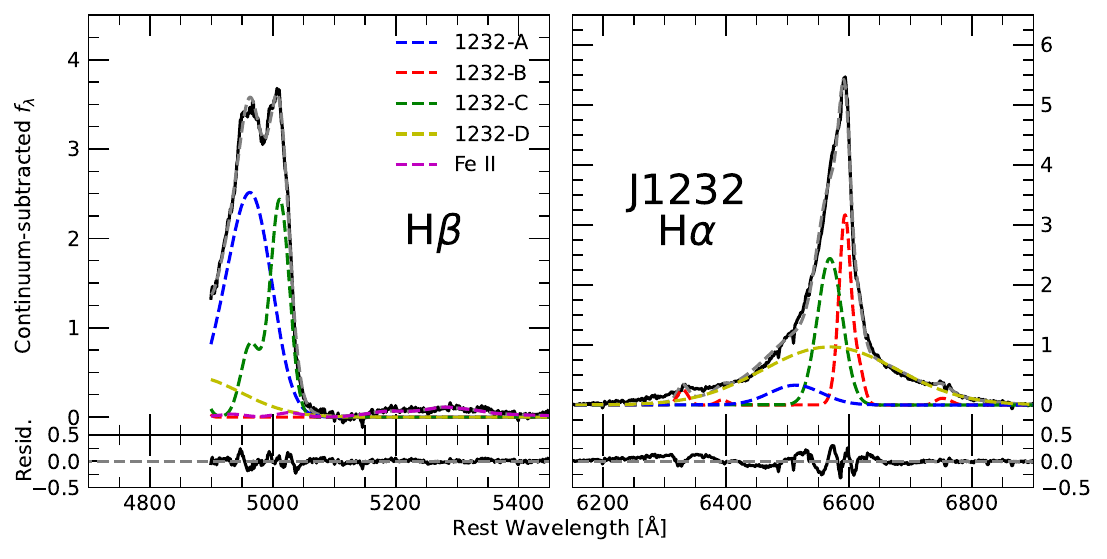}}
\subfigure{
\includegraphics[width=0.45\textwidth]{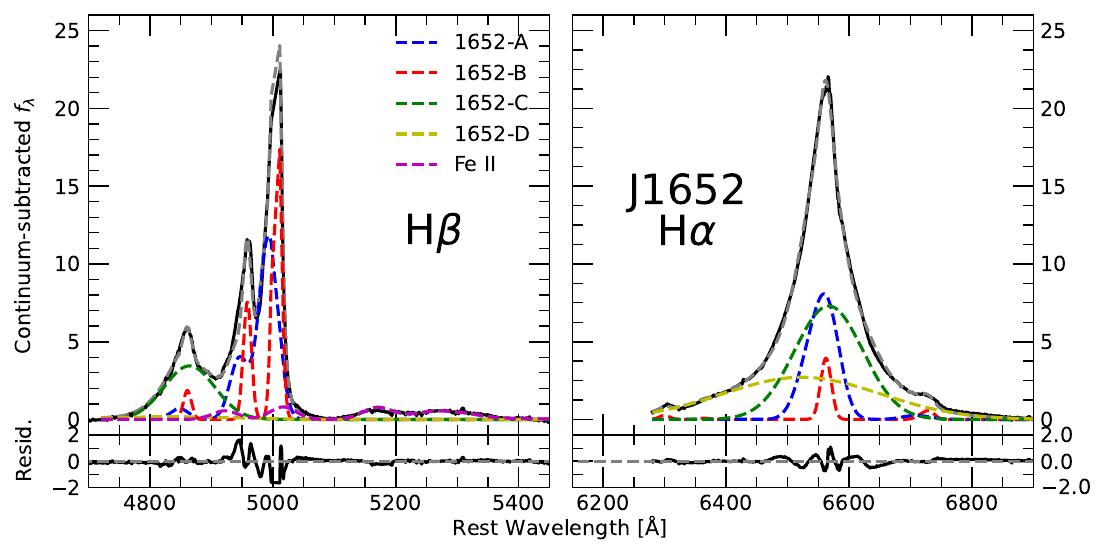}}
\hfill
\subfigure{
\includegraphics[width=0.45\textwidth]{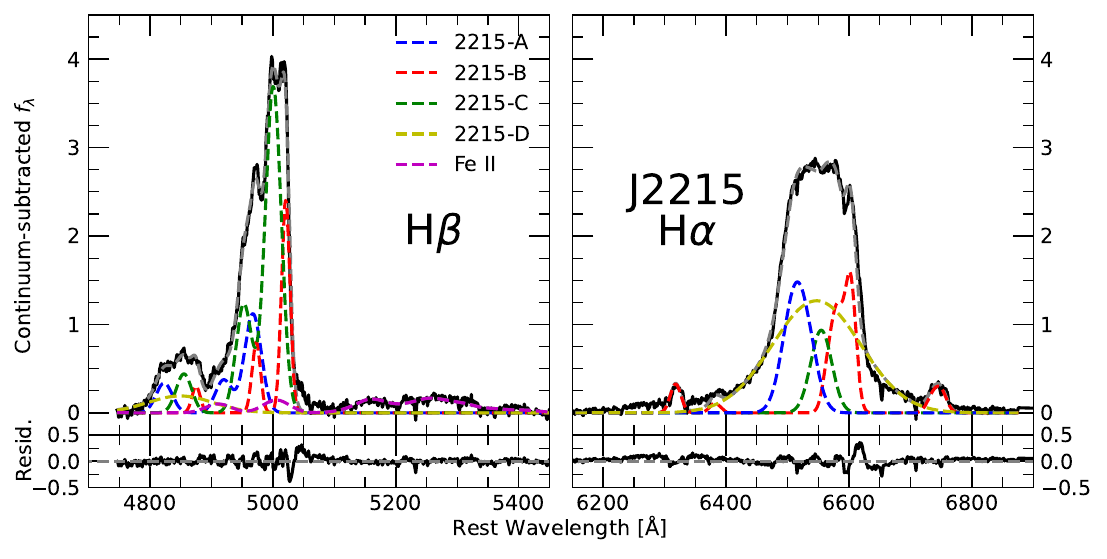}}
\hfill
\caption{Modeled emission profiles for the Balmer profiles of J0832, J1217, J1232, J1652, and J2215.  The blue/red/green/yellow lines correspond to the ``A''/``B''/``C''/``D'' templates in Table~\ref{tab:erq_templates}, respectively.}\label{fig:erqs-balmer}
\end{sidewaysfigure*}

\subsection{J0832}

We use 0832-A, -B, and -D to fit the emission lines around \halpha\ and \hbeta.  There is no evidence for additional blueshifted emission beyond that modeled by Template 0832-A, so we do not include a template ``C.''  The blue wing of the \hbeta\ emission is cut off by the NRS1 edge, so all of our constraints on the \hbeta\ emission (except for perhaps 0832-A) are from the \halpha.  However, when we use Templates 0832-A and -C to predict the full profile of the \hbeta\ emission, we find that it matches the line profile in the X-Shooter spectrum, as seen in Figure~\ref{fig:others-hbeta}.

Template 0832-B fits the narrowest \halpha\ and \hbeta\ emission, and it is the only template that includes [\ion{O}{1}] and [\ion{S}{2}].  Meanwhile, the \othree\ emission appears to be produced exclusively by Template 0832-A. 0832-A is blueshifted relative to 0832-B by $\sim$2000\kms.  Both templates include moderate [\ion{N}{2}] emission.  The broadest \halpha\ emission is well fit by 0832-D, and we appropriately fit the \ion{Fe}{2} emission with a single template (unlike J0834 which needs two). 

Template 0832-B likely corresponds to NLR-like gas associated with the galaxy, whereas 0832-A shows clear outflow signatures with very high \othree\ velocities, uniquely characteristic of the ERQ population.  0832-D corresponds to the BLR of the central source.  

\subsection{J1217}

We use four templates to fit the emission around \halpha\ and \hbeta.  The blue wing of the \hbeta\ emission is again cut off by the NRS1 edge, so the constraints on the \hbeta\ emission are mostly due to the fits to \halpha.  Template 1217-B, which produces the [\ion{O}{1}], [\ion{S}{2}], and a significant portion of the \othree\ emission, has a FWHM of 650\kms.  This is the narrowest of all the templates included in Tables~\ref{tab:j0834} and \ref{tab:erq_templates}.  There is an \othree\ emission component in Template 1217-A that is marginally offset relative to 1217-B, and indeed 1217-A may correspond to a higher-velocity component of the gas associated with 1217-B.  1217-C and 1217-D produce the bulk of the \halpha\ and \hbeta emission which are dominated by BLR-like gas.  The presence of obvious narrow lines makes J1217 closer to a very-reddened version the ``typical'' AGN compared to the other ERQs.  This agrees with the SED models of J1217 in Section~\ref{sec:sed}.

\subsection{J1232}

\begin{figure}
\centering
\centering
\subfigure{
\includegraphics[width=0.45\textwidth]{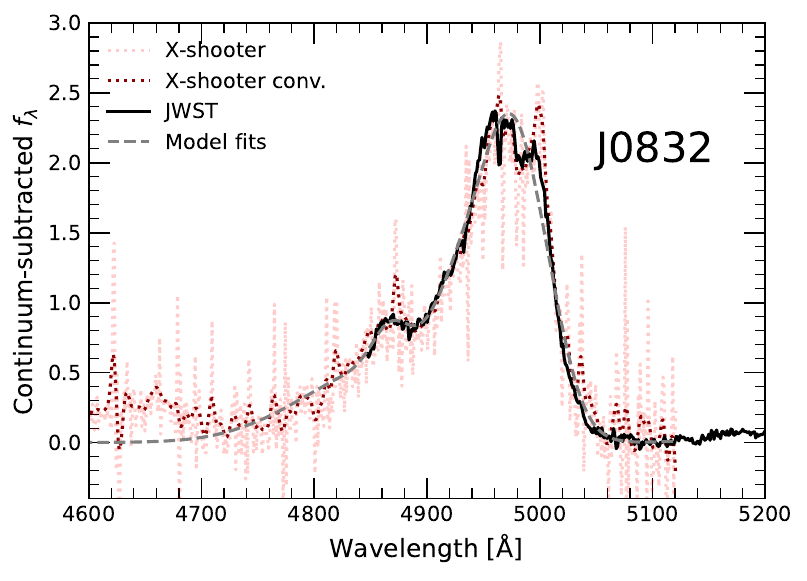}}
\subfigure{
\includegraphics[width=0.45\textwidth]{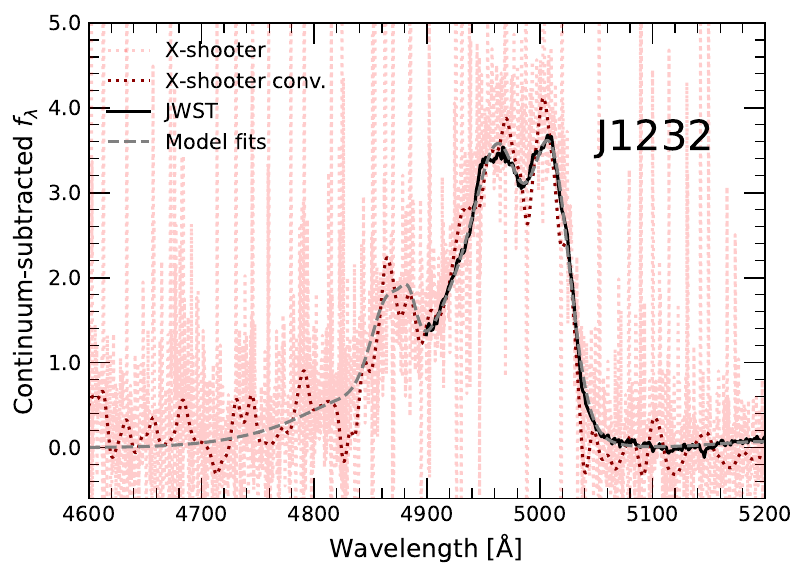}}
\caption{Model fits to the \hbeta\ emission for J0832 (\textbf{top}) and J1232 (\textbf{bottom}).  The JWST spectra are shown in black, the X-Shooter spectra is shown in red, and smoothed versions of the X-Shooter spectra are shown in dark red.  Our model fits where the blue wing is constrained by the fit to \halpha\ are  shown in gray. Our extrapolated models are able to capture the \hbeta\ emission of the X-Shooter spectrum, despite having no information of emission blueward of $\sim$4900\ang. The Balmer decrements for several of the templates (0832-A, 1232-A, 1232-B, and 1232-C) are unconstrained and effectively set at the lower limit of 2.86.  }
\label{fig:others-hbeta}
\end{figure}

We use four templates to fit the emission around \halpha\ and \hbeta.   The \hbeta\ emission is entirely cut off by the NRS1 edge.  Of particular note is the peculiar double-peaked \othree\ emission.  The emission is due to two components with a velocity offset such that the~\lam5007 line of one component (represented by 1232-A) is nearly coincident with the~\lam4959 line of the other (1232-C).  Interestingly, there is no strong evidence for \othree\ emission in Template 1232-B, which is associated with the reddest velocity shift and the narrowest emission of \halpha.  The component also includes [\ion{O}{1}], [\ion{S}{2}], and a small amount of [\ion{N}{2}].  The remaining very-broad \halpha\ emission is modeled by 1232-D.   In Figure~\ref{fig:others-hbeta}, we show that our JWST-derived model of the \hbeta\ emission fits the \hbeta\ emission profile of the X-Shooter spectrum reasonably well.


Like with the other ERQs in our sample, 1232-B seems to trace emission associated with NLR-like gas of the host galaxy, and 1232-D matches what we would expect from a BLR.  1232-A and 1232-C are more difficult to assess, as they model the broad, double-peaked \othree\ (and some \halpha). Other AGNs with double-peaked \othree\ profiles that are thought to be either dual-AGNs with kinematically separated NLRs (e.g., \citealt{liu10,liu18b,yuan16}) or biconical outflows (e.g., \citealt{greene12,liug13a,liug13b}).  The latter population is thought to be the majority of sources with double-peaked \othree. 

\subsection{J1652}

We use four templates to fit the emission around \halpha\ and \hbeta.  We have the full profile of \hbeta\ emission for this source.  We can compare our model of the emission line profiles of J1652 with those of \citet{bertemes25}.  For the most part they agree in the modeled velocity offsets and dispersions.  There are some small discrepancies, but this can be largely attributed to the differences in the modeling approach. For example, \citet{bertemes25} did not require each component of their fits to the \halpha\ and \hbeta\ profiles to be kinematically linked. 

Templates 1652-A, -B, -C are all best fit with Balmer decrements equal to 2.86, implying that \textit{all} of the optical emission line flux is either from scattered light or from gas on scales larger than the continuum-obscuring gas.  

\subsection{J2215}

We use four templates to fit the emission around \halpha\ and \hbeta.  We have the full profile of \hbeta\ region for this source.  The most intriguing feature of spectra is the flat-topped \halpha\ and \hbeta\ emission profiles.  We are able to model them as a combination of the four templates, as seen in Figure~\ref{fig:erqs-balmer}, but, due to significant blending and lack of individual sharp features indicative of a given template, this may not accurately capture the physical conditions that create these emission profiles.  Like with J1232, we believe this emission, along with the peculiar \othree\ emission profile, is indicative of biconical outflows.  \\

\section{Continuum Emission Fitting}\label{sec:sed}

\subsection{UV/optical/IR SED Fitting}

We are interested in the SEDs of our ERQ sample because it is clear they do not resemble the typical AGN.  \citet{alexandroff18} find that the rest-frame UV continuum and emission lines of ERQs are highly polarized, so there is strong evidence that these wavelengths and perhaps the rest-frame optical are dominated by scattered light rather than by direct emission.  We adopt this physical intuition in our model.  Our model mainly consists of two templates: (1) the QSO1 template from the SWIRE Template Library \citep{polletta07} for the scattered light; and (2) the torus templates from \citet{martinez24} based on \citet{stalevski16} for the IR continuum from the IR-bright torus.  The \citet{stalevski16} templates are a function of a number of parameters, and \citet{martinez24} averages these models and presents them as function of inclination angle $\theta$, and so we use this framework to allow for flexibility in modeling the torus.  We do not include a host galaxy template as the host galaxy luminosities are several orders of magnitude below the observed emission \citep{zakamska19,wang25,chenyc25}.
 
In constructing the observed SEDs, we use SDSS photometry and spectra, the X-Shooter spectra, JWST spectra, and the WISE photometry for each ERQ.  For the SDSS, X-Shooter, and JWST spectra, we synthesize the spectra into artificial photometry bins that mimic the overall shape of the continuum emission, taking the median flux between 1050--1150\ang, 1250--1450\ang, 5250--5750\ang, and 7250--7750\ang.  We do this to avoid contamination from strong emission and absorption lines.  For ERQs with X-Shooter spectra, we scale the SDSS and 

We use a torus template $T(\theta)$ which is a function of the inclination angle $\theta$, and a scattered light template $S$ which is extincted by $A_V$ so that the overall SED model at a given wavelength $\lambda$ is 
$$\lambda f_\lambda = T_\lambda(\theta) + f_{\rm sc} \cdot S_\lambda \cdot 10^{-A_\lambda/2.5}$$ 
where $f_{\rm sc}$ is the scattered light fraction at W4 and $A_\lambda$ is the extinction at wavelength $\lambda$.  We optimize the fit on a grid of $\theta$, $f_{\rm sc}$, and $A_V$, setting the normalization factor so that the model matches the W4 flux.  The fit is evaluated using the root-mean-squared errors between the model and the observed logarithmic fluxes.

The values for $f_{\rm sc}$ and $A_V$ are dependent on the extinction curve and $R_V$ used.  Using an extinction law that is flatter leads to higher values of $A_V$ and $f_{\rm sc}$ -- e.g., the \citet{gaskell07} curve that has significantly less extinction in bluer wavelengths compared to the \citet{cardelli89} curve at fixed $A_V$.  The same effect occurs when using a higher $R_V$.  While the different extinction curves has implications for the dust grain size distribution of the obscuring medium, this topic is not the focus of this paper, and so we adopt a \citet{calzetti00} extinction curve with $R_V = 4.0$ which, roughly, lands in between the range of extinction values for \citet{cardelli89} and \citet{gaskell07} curves with $R_V = 3.1$--5.

\begin{figure*}
\includegraphics[width=\linewidth]{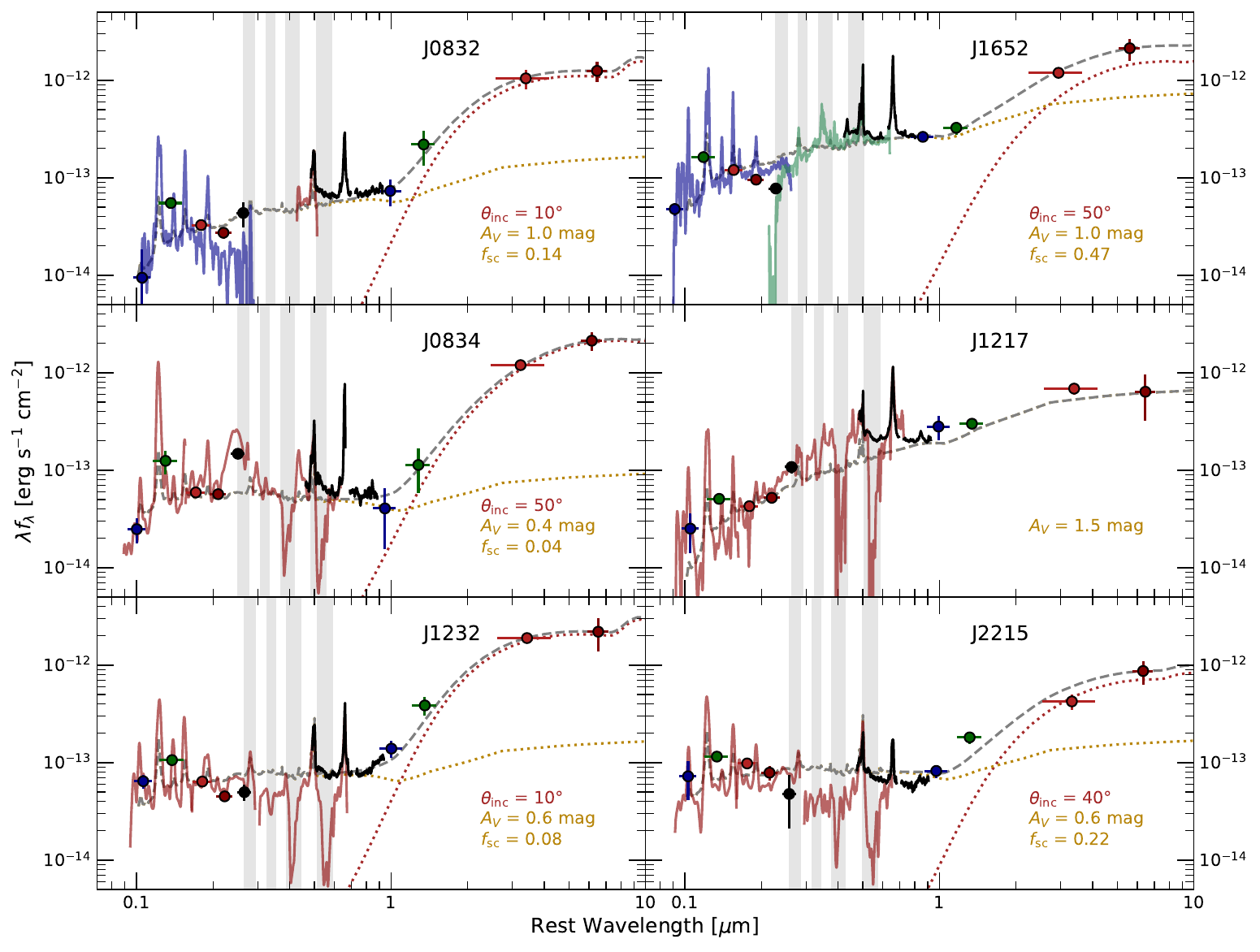}
\caption{UV to MIR SEDs of our ERQs.  The solid lines are the JWST (black) and archival spectra (X-Shooter - dark red, SDSS - dark blue, GNIRS - green), and the solid colored circles are the archival SDSS and WISE photometry.  Atmospheric absorption bands are shaded gray.  The dotted brown line is the best-fit torus template, the dotted gold line is the best-fit scattered light template, and the dashed grey line corresponds to the combined model.  The physical parameters of each model are included in the lower-right corner of each panel.  Because J1217 is best fit with a simple reddened QSO1 template, we only give the $A_V$.}
\label{fig:sed}
\end{figure*}

The best-fit SED models are shown in Figure~\ref{fig:sed}.  For all of the ERQs except J1217, the SED is fit by the torus + scattered light model with moderate extinction $A_V < 1.5$\,mag.  For J0834, the SED fit yields $A_V=0.4$\,mag, which is comparable to the lowest Balmer-decrement-derived $A_V$ of Template 0834-D of $A_V =0.4$\,mag and the \ion{O}{1}~\lam1304/\lam8446 ratio-derived $A_V = 0.5$\,mag.   While the shape of optical \textit{continuum} flux of J0834 requires it to be mostly from a scattered component, the optical \textit{emission line} flux ought to mostly be due to line-of-sight obscuration as implied by the large extinction coefficients derived from the Balmer decrements.  
 
J1217 is an interesting exception, in that the SED can be modeled entirely with the scattered light template.  Rather than this meaning the emission from J1217 is entirely scattered, the most likely explanation is that the emission we see is directly-observed and reddened, rather than being a combination of a directly-observed torus and scattered light. This is matched by the relatively smooth shape of the SED from 1000\ang\ to 10\mum.  The fact that J1217 appears to be a more similar to a ``typical'' Type 1 AGN (compared to other ERQs) matches with our analysis in Section~\ref{sec:spec} in that J1217 has relatively narrow forbidden lines, again more comparable to a typical Type 1 AGN than the broad forbidden lines of the other ERQs.   

The scattering fractions ($f_{\rm sc}$) we fit and present in Figure~\ref{fig:sed} are based on the fraction of W4 produced by the scattered light template.  While these values are not excessively large, when we extrapolate down to the UV and optical range, our model predicts that most if not all of the continuum flux is scattered light.  If we take J0834 as an example, our model predicts that only $\sim$4\% of the emitted UV/optical flux is scattered, but that \textit{all} of the \textit{observed} UV flux is scattered light.  This is analogous to Type 2 AGNs with ``hidden'' polarized blue continua and broad emission lines that must be from scattering, where the polarization fraction is only $\sim$a few percent \citep{miller90,antonucci93,zakamska05}.  ERQs, meanwhile, have polarization fractions of up to 20$\%$ \citep{alexandroff18}.  

\subsection{Sub-mm and Radio Emission}
 
The decomposition between the warm dust emission from the AGN and the cool dust emission from the host galaxy is very uncertain, and some empirical torus and AGN templates contain excess flux at FIR and sub-mm wavelengths that could well be due to the under-subtracted host galaxy.  The level of this emission could very well be dependent on the luminosities of the original objects used in template construction. Without additional probes of star formation in the host galaxy, it is not easy to fully disentangle these two sources of emission. For the same MIR flux normalization, the sample models included in \citet{martinez24} show a several dex range in FIR flux.

\begin{figure*}
\includegraphics[width=\linewidth]{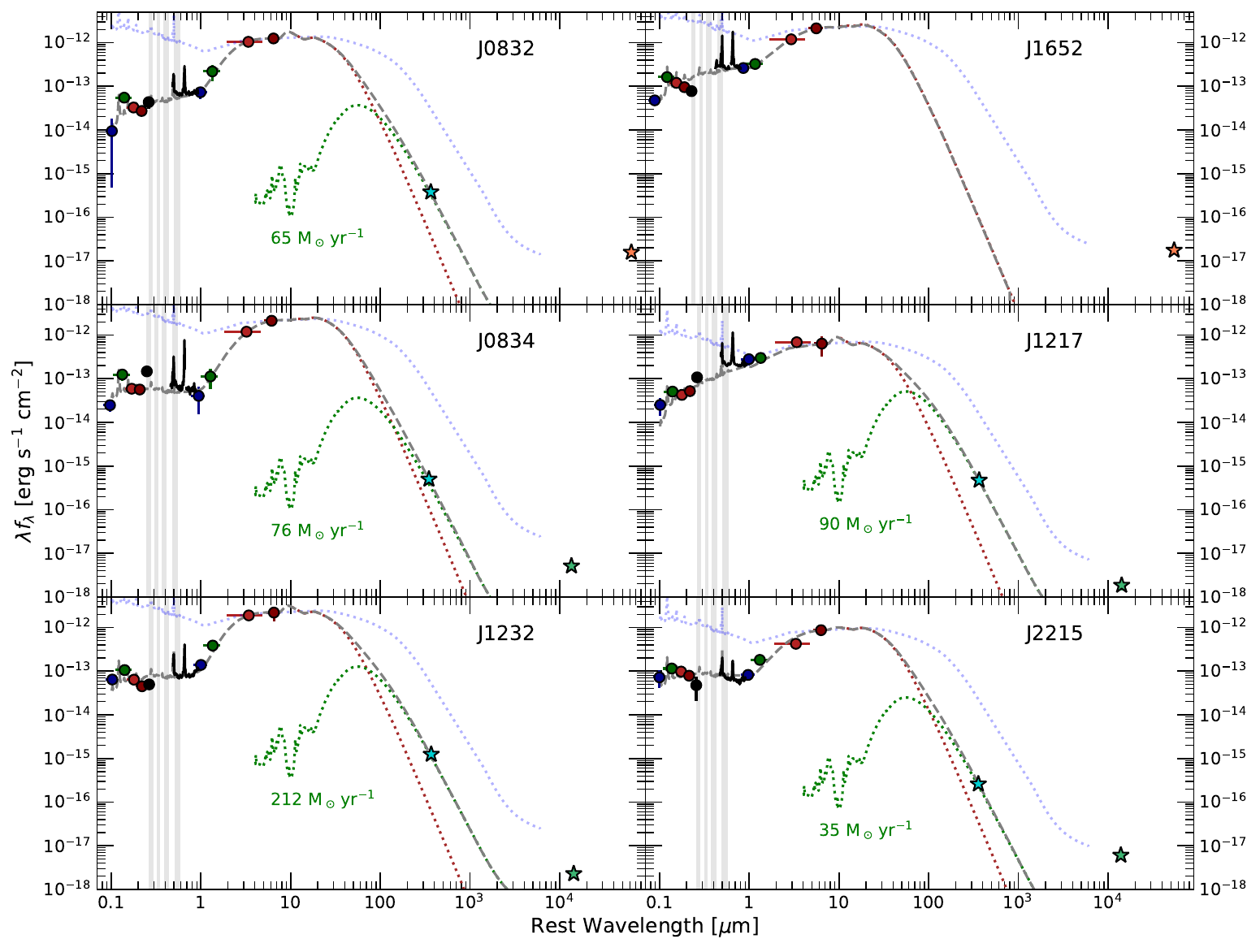}
\caption{Full SED of our ERQs.  The UV to MIR SED models are the same as in Sec.~\ref{sec:sed}.  The predicted torus template flux is plotted as a brown dotted line.  The QSO1 template is plotted as a blue dotted line.  The ALMA, VLA, and FIRST fluxes are plotted as blue, red, and green stars, respectively.  The \citet{rieke09} SF SED is plotted as a dotted green line, and it is scaled to match the difference between the observed ALMA flux and the predicted torus template flux.  The corresponding SFR is included. }
\label{fig:alma}
\end{figure*}

As seen in Figure~\ref{fig:alma}, when we scale the scattered light QSO1 SWIRE template to WISE fluxes, the predicted flux in the ALMA band is much higher than what is observed.  It is likely that this model overpredicts flux beyond 20\mum\ -- indeed, the \citet{stalevski16} models are much fainter than the SWIRE template at this wavelength.  To adapt a model for the full SEDs of these ERQs, we split it into two.  Blueward of W4 band, we adopt the SED fits as above (a composite of scattered light and torus templates), but redward of the W4, we treat it as coming entirely from the torus template component.  

Using this approach, we find that our combined models actually underpredict the observed ALMA fluxes.  It is possible that this additional observed flux is coming from star formation (SF) in the host galaxy.  To model the excess, we use the \citet{rieke09} SF SED template, scaling it such that the combined SF and torus templates match the observed ALMA flux.  Based on this scaling, we compute the total IR fluxes and convert them to SF rates (SFRs) using \citet{bell03}.  The resulting SFRs (30--200\,M$_\odot$\,yr$^{-1}$) are reasonable compared to what is found for other studies of this population of ERQs \citep{chenyc25} and for galaxies with these masses at cosmic noon \citep{looser25}.  Again, these measurements are a strict upper limit on the SFR, as the AGN may indeed be contributing more flux in the FIR/sub-mm depending on the details of the geometric distribution of the dust.

When we include the VLA observations of our ERQs from \citet{hwang18}, we can see that the observed radio fluxes are higher than what is expected from SF, and so there must be some amount of radio emission from the quasars.  This is not unexpected, as we can see that the QSO1 SWIRE template includes flux at these wavelengths.  The origin of this radio emission in otherwise nominally radio-quiet quasars is not well understood, although it is well established that in these extremely high-luminosity quasars SF is insufficient \citep{zakamska16,hwang18}. The radio emission can be due to weak jets, although there is no evidence for that in the radio morphology or the spectral indices in the ERQ population \citep{hwang18}, or it can be the by-product of quasar-driven winds which are extremely powerful in this population \citep{zakamska14,hwang18}. \\

\section{Discussion} \label{sec:discuss}

Based on our results of Sections~\ref{sec:j0834}, \ref{sec:spec}, and~\ref{sec:sed}, the overall apparent fluxes of our ERQs (with the exception of J1217) are best understood as being dominated by emission obscured by dusty gas along our line-of-sight combined with emission scattered into our line-of-sight by a relatively-dust-free gas.  The UV-optical continuum as well as the UV emission lines are due to scattered light, while the red optical-NIR continuum are obscured and reddened along the line-of-sight.  The optical emission lines must be a combination of directly-observed/obscured and scattered emission, consistent with range of observed Balmer decrements.

\begin{figure}
\includegraphics[width=0.99\linewidth]{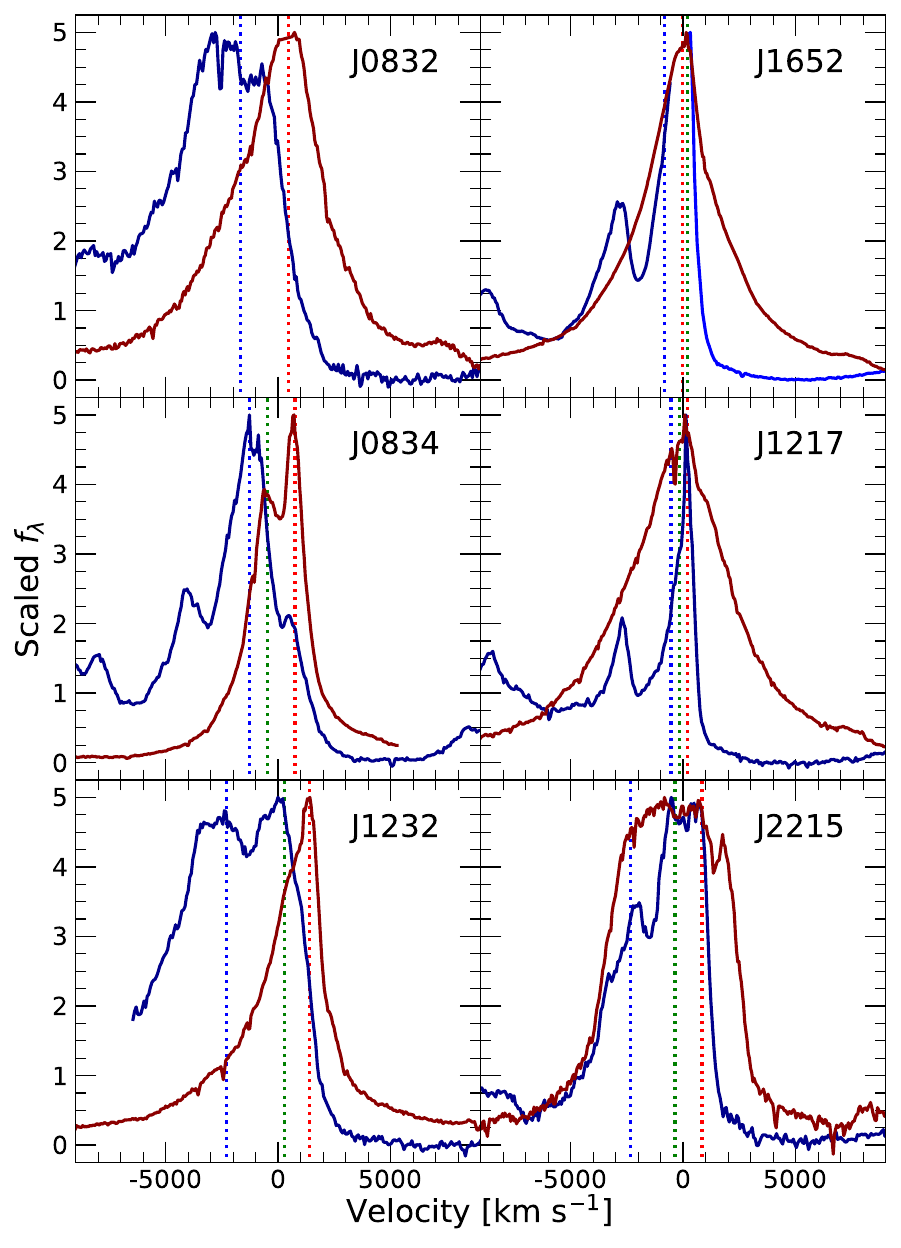}
\caption{Velocity profiles of the \halpha\ (dark red) and \othree\ (dark blue) emission for each ERQ.  We scale the \othree\ and \halpha\ to each other for better visual comparison.  We also plot the centroids of our model templates from Tab.~\ref{tab:j0834} and \ref{tab:erq_templates} as colored dotted lines, where the colors are the same as Figs.~\ref{fig:j0834-balmer} and \ref{fig:erqs-balmer}.}
\label{fig:hao3}
\end{figure}

Using our spectral templates in Sections~\ref{sec:j0834} and \ref{sec:spec}, we match the emission components that are likely produced by different physical processes on different spatial scales.  We summarize some of our key findings in Figure~\ref{fig:hao3}, which highlights the differences in kinematic structure between the \halpha\ and \othree\ emission.   Based on a combination of velocity offsets, dispersions, and emission line ratios, the templates can roughly be divided into distinct physical sources:

\textit{High-velocity \othree-dominated outflows:} 0832-A, 0834-A+C, 1217-A, 1232-A+C, 2215-A+C 

\textit{NLR-like gas:} 0832-B, 0834-B, 1217-B, 1232-B, 1652-B(+A?), 2215-B

\textit{BLR-like gas:} 0832-D, 0834-D, 1217-C+D, 1232-D, 1652-D(+C?), 2215-D

Since the emission that would be normally attributed to just a ``typical'' AGN (NLR+BLR) is supplemented by high-velocity \othree-dominated outflows, this explains the relatively high EWs as measured in \citet{hamann17} -- in other words, the emission line flux relative to the continuum is artificially inflated by substantial emission from gas that is not ``part'' of the central AGN. Moreover, because the continuum flux must mostly be scattered rather than obscured emission along the line of sight (as implied by the low $A_V$ values from SED modeling), variations in the amount of scattered continuum light and the amount of obscuration (or lack thereof) of the emission-line-dominated outflows will change the relevant EWs.  This is suggested in \citet{alexandroff18}, where the continuum in ERQs is typically more polarized than the (UV) emission lines.

\subsection{J0834}

As in Section~\ref{sec:j0834}, we highlight our discussion of J0834.  Combining our spectral analysis with our SED analysis in Section~\ref{sec:sed}, we can create a composite picture of the physical systems producing our observations.  The multiple templates modeled to the spectral features of J0834 can be understood as follows: 0834-A and 0834-C are correspond to outflows, and 0834-B and 0834-D are features intrinsic to the central AGN itself: 0834-B corresponding to a NLR and 0834-D corresponding to a BLR.

Most of the \othree\ is produced by 0834-A, and the majority of the Balmer lines are produced by 0834-C.  Put another way, the \othree\ emission is systematically blueshifted relative to the Balmer lines. The relative ratios of 0834-A to 0834-C for other lines varies (see Figs.~\ref{fig:j0834-other} and \ref{fig:j0834-oi}). 0834-A and 0834-C may represent two stages of the same wind, as the templates' velocity offsets are comparable to the FWHMs.  The difference between the stages could be due to complicated temperature or density gradients or ionization physics.  For example, the strong \ion{O}{1}~\lam1304 and~\lam8446 implies Ly$\beta$ fluorescence in optically thick gas, which we would expect to suppress \othree\ by approaching its critical density.  This could in part explain why the bulk of the \othree\ is blueshifted relative to the bulk Balmer emission; the \othree\ is being produced by a less-dense, higher-velocity component -- or region of the outflow -- than other emission. 

We show that the \ion{O}{1}~\lam1304 and~\lam8446 and the Balmer lines are kinematically related.  Yet, the flux ratios of the \ion{O}{1} show evidence for relatively little obscuration, whereas the Balmer decrements imply much more obscuration.  This means that the UV \ion{O}{1} line and, likely, the entirety of the UV flux, is scattered into our line-of-sight rather than directly observed, in line with the findings of \citet{alexandroff18}.  This evidence is supplemented in the form of excess UV \ion{Fe}{2}, which is indicative of fluorescence from UV-bright continuum \citep{wang16}.  Moreover, the low extinction value implied by the \ion{O}{1} line ratio, combined with the fact that Ly$\beta$ fluorescence traces dense gas, means that the emitting gas -- the outflows and the NLR gas -- must be relatively dust-free.  

The $A_V$ values implied by the Balmer decrements of Templates 0834-B and 0834-C are not dramatically different from what would redden a typical mid-luminosity AGN SED to the observed optical-to-IR SED.  It would, however, require a higher obscuration value of $A_V > 8$ to match the shape of the NIR SED -- more specifically, the WISE colors -- though this can be explained by variations in the NIR SED due to different potential torus structures \citep{stalevski16,martinez24}.  Overall, it is likely that the optical emission line flux we see is a combination of both scattered and directly-observed emission, with the former being less reddened and the latter being significantly reddened by dust, the result being a ``combined'' Balmer decrement.  

Because Template 0834-A, which models the bulk of the \othree\ emission, is much less reddened compared to the 0834-B and 0834-C, it is likely that its associated outflow is on larger scales than the source of the obscuration.  The small critical density of \othree\ also implies that the outflow is on large scales where the gas is more rarefied.  As stated earlier, 0834-A and 0834-C may be stratified components of the same outflow, where 0834-A is the larger-scale, high-velocity, and unobscured component and 0834-C is the more compact, low-velocity, and obscured component.

The location of the scattering gas relative to the outflow(s) is uncertain, but because the \ion{O}{1}~\lam1304 emission, which we know must be scattered, is not dominated by 0834-A in the way the \othree\ emission is, the outflow is on similar or larger scales than the scattering gas.  Is it possible that the scattering gas is the same as the outflowing gas?  We know that the continuum and emission lines of ERQs are spatially compact \citep{lau24}, and therefore the spatial scales of the scattering gas and the outflowing gas are likely not dramatically different (or at least, they are both sufficiently compact).  

However, this is where the velocity differences between the scattered and obscured emission would become problematic.  If the outflowing gas was the scattering gas, the scattered emission would be systematically redshifted relative to the line-of-sight obscured emission \citet{zakamska23}.  Yet, we know that there are not significant differences between the kinematics of the Balmer lines, which we know to be obscured, and the kinematics of the \ion{O}{1}~\lam{1304} line, which must be entirely scattered.  It is therefore likely that the scattering gas and the outflowing gas are not the same.  

\subsection{The other ERQs}

There is a clear diversity of kinematic structures among our sample of ERQs, and so it is likely that the physical scenario of J0834 derived from its observations is not the exact same physical scenarios of the other ERQs.  

We note that there is no evidence for strong UV \ion{Fe}{2} associated with the other ERQs in comparison to J0834, which we link to evidence for fluorescent scattering.  We are unsure of what this means for the other ERQs, but it may be linked to the fact that J0834 has more distinct velocity offsets, e.g., the clearly double-peaked profiles in \halpha, \hbeta, and \othree, in its spectral features compared to the other ERQs. \\

\section{Summary}\label{sec:conc}

Our findings can be summarized:
    
\begin{enumerate}[leftmargin=*]
\item[$\bullet$] We analyze the optical spectra of a sample of ERQs observed by JWST using multi-component spectral templates.  These templates allow us to split the emission into different kinematic components that are coming from BLR-like clouds, from low-density NLR-like gas, and from high-velocity, low-density outflows on scales larger than the obscuring gas.
\item[$\bullet$] We combine our JWST observations with archival data to analyze the SEDs of these ERQs and find they are consistent with a significantly dust-obscured central source with a small amount of relatively-unobscured UV/optical flux that is scattered into our line-of-sight by gas that is also on larger scales than the obscuring gas.
\item[$\bullet$] We highlight in particular J0834 for its distinct emission line kinematics.  Combining extinction estimates from the Balmer decrement and \ion{O}{1} line ratios, we show that, while the kinematics of the UV and optical emission lines are very similar, the observed UV line fluxes must be entirely scattered light while the optical emission fluxes are dominated by directly-observed and reddened emission.  Moreover, the properties of the \ion{O}{1} emission imply that the emitting gas, including the outflows, must be relatively dust-free.
\end{enumerate}

Our work highlights the kinematic structures of outflows in the central regions of some of the most luminous objects in the Universe.  Future work will include analysis of the kinematics of outflows on different scales: namely, the cold molecular gas of the ERQ host galaxies and the shocked gas around these ERQs (Vayner et al. in prep).  In furthering our understanding of the outflows of these luminous and obscured AGNs, we also further our understanding of the complex evolutionary dynamics between AGNs and their host galaxies. \\

\section*{Acknowledgements}

JMMN thanks CS Kochanek for helpful comments and suggested edits.  JMMN was supported by a William H.~Miller III Fellowship.  NLZ thanks the Institute for Advanced Study in Princeton, NJ for hospitality during regular visits. SV was supported in part by NASA through STScI grant JWST-ERS-01335.

This work is based  on observations made with the NASA/ESA/CSA James Webb Space Telescope. The data were obtained from the Mikulski Archive for Space Telescopes at the Space Telescope Science Institute, which is operated by the Association of Universities for Research in Astronomy, Inc., under NASA contract NAS 5-03127 for JWST. These observations are associated with program \#1335 and \#2457.

This paper makes use of the following ALMA data: ADS/JAO.ALMA\#2017.1.00478.S. ALMA is a partnership of ESO (representing its member states), NSF (USA) and NINS (Japan), together with NRC (Canada), NSTC and ASIAA (Taiwan), and KASI (Republic of Korea), in cooperation with the Republic of Chile. The Joint ALMA Observatory is operated by ESO, AUI/NRAO and NAOJ.

\bibliographystyle{mnras}
\bibliography{bibliography}


\end{document}